\documentclass[aps,superscriptaddress,amsmath,amssymb,floatfix,twocolumn,showpacs,amsfonts,longbibliography]{revtex4-1}
\usepackage{times}
\usepackage[varg]{txfonts}
\usepackage{textcomp}
\usepackage{graphicx}
\usepackage{subfigure}
\usepackage{color}
\usepackage[colorlinks=true,citecolor=blue,urlcolor=blue,linkcolor=blue,hyperindex]{hyperref}
\usepackage{braket}
\usepackage{overpic}

\def\pp{\mathbf{p}}
\def\kk{\mathbf{k}}
\def\qq{\mathbf{q}}
\def\bs{\mathbf{S}}

\def\bq{\mathbf{Q}}
\def\bdelta{\boldsymbol{\delta}}
\def\non{\nonumber}

\begin{document}

%\title{Probing magnetic excitations with indirect resonant inelastic x-ray scattering in spiral antiferromagnets on the triangular lattice}
\title{Torque equilibrium spin wave theory study of anisotropy and Dzyaloshinskii-Moriya interaction effects on the indirect K$-$ edge RIXS spectrum of a triangular lattice antiferromagnet}
\author{Shangjian Jin}
\affiliation{State Key Laboratory of Optoelectronic Materials and Technologies, School of Physics, Sun Yat-Sen University, Guangzhou 510275, China}

\author{Cheng Luo}
\affiliation{State Key Laboratory of Optoelectronic Materials and Technologies, School of Physics, Sun Yat-Sen University, Guangzhou 510275, China}

\author{Trinanjan Datta}
\email[Corresponding author:]{tdatta@augusta.edu}
\affiliation{State Key Laboratory of Optoelectronic Materials and Technologies, School of Physics, Sun Yat-Sen University, Guangzhou 510275, China}
\affiliation{Department of Chemistry and Physics, Augusta University, 1120 15th Street, Augusta, Georgia 30912, USA}

\author{Dao-Xin Yao}
\email[Corresponding author:]{yaodaox@mail.sysu.edu.cn}
\affiliation{State Key Laboratory of Optoelectronic Materials and Technologies, School of Physics, Sun Yat-Sen University, Guangzhou 510275, China}

\date{\today}

\begin{abstract}
We apply the recently formulated torque equilibrium spin wave theory (TESWT) to compute the $1/S$-order interacting $K$ -edge bimagnon resonant inelastic x-ray scattering (RIXS) spectra of an anisotropic triangular lattice antiferromagnet with Dzyaloshinskii-Moriya (DM) interaction. We extend the interacting torque equilibrium formalism, incorporating the effects of DM interaction, to appropriately account for the zero-point quantum fluctuation that manifests as the emergence of spin Casimir effect in a noncollinear spin spiral state. Using inelastic neutron scattering data from Cs$_2$CuCl$_4$ we fit the 1/S corrected TESWT dispersion to extract exchange and DM interaction parameters. We use these new fit coefficients alongside other relevant
model parameters to investigate, compare, and contrast the effects of spatial anisotropy and DM interaction on the RIXS spectra at
various points across the Brillouin zone. We highlight the key features of the bi- and trimagnon RIXS spectrum at the two inequivalent rotonlike
points, $M(0,2 \pi/\sqrt{3})$ and $M^{\prime}(\pi,\pi/\sqrt{3})$, whose behavior is quite different from an isotropic triangular
lattice system. While the roton RIXS spectrum at the $M$ point undergoes a spectral downshift with increasing anisotropy, the peak at the
$M^\prime$ location loses its spectral strength without any shift. With the inclusion of DM interaction the spiral phase is more stable and the peak at both $M$ and $M^\prime$ point exhibits a spectral upshift. Our calculation offers a practical example of how to calculate interacting RIXS spectra in a non-collinear quantum magnet using TESWT. Our findings provide an opportunity to experimentally test the predictions of interacting TESWT formalism using RIXS, a spectroscopic method currently in vogue.
\begin{description}
\item[PACS number(s)] 78.70.Ck, 75.25.-J, 75.10.Jm
\end{description}
\end{abstract}

\maketitle

%%%%%%%%%%%%%%%%%%%%%%%%%%%%%%%%%%%%%%%%%%%%%%%%%%%%%%%%%%%%%%%%
\section{Introduction}\label{sec:intro}
%%%%%%%%%%%%%%%%%%%%%%%%%%%%%%%%%%%%%%%%%%%%%%%%%%%%%%%%%%%%%%%%
In a recent publication Cheng~\emph{et. al.}, Ref.~\onlinecite{PhysRevB.92.035109}, highlighted the features of the indirect $K$ -edge resonant inelastic x$-$ray scattering (RIXS) bi- and trimagnon spectrum of an isotropic triangular lattice antiferromagnet (TLAF).  The TLAF is known to possess a $120^\circ$ long range ordered state even after quantum fluctuations are considered \cite{SinghPhysRevLett.68.1766,HusePhysRevLett.60.2531,PhysRevB.50.10048,PhysRevB.91.014426,PhysRevB.67.104431,KadowakiJPSJ,IshiiEPL,PhysRevB.81.104411,TothPhysRevB.84.054452,TothPhysRevLett.109.127203,
PhysRevLett.108.057205,PhysRevB.91.024410,PhysRevLett.110.267201}. The authors considered the self-energy corrections to the spin-wave spectrum to pinpoint the nontrivial effects of magnon damping and very weak spatial anisotropy on RIXS. It was shown that for a purely isotropic TLAF model, a multipeak RIXS spectrum appears which is primarily guided by the damping of the magnon modes. Interestingly enough it was demonstrated that the roton momentum point is immune to magnon damping (for the isotropic case) with the appearance of a single-peak RIXS spectrum. It was suggested that this feature could be utilized as an experimental signature to search for or detect the presence of roton like excitations in the lattice. However, including XXZ anisotropy leads to additional peak splitting, including at the roton wave vector.

At present no theoretical guidance exists for experimentalists on how to interpret the RIXS spectrum of the ordered phase in a geometrically frustrated triangular lattice quantum magnet, though a proposal has been put forward to detect spin-chirality terms in triangular-lattice Mott insulators via RIXS~\cite{PhysRevB.84.125102}. Furthermore, as discussed in this article the existing spin wave theory formulation used for the isotropic case fails beyond the isotropic point and with Dzyaloshinskii-Moriya (DM) interaction included in the model.

Lately, the nature of the ground and excited states of the TLAF has garnered some attention~\cite{ChenPhysRevB.87.165123,SchmidtPhysRevB.89.184402,hauke2011modified,nphys749,fishmanPhysRevB.79.184413,PhysRevB.81.020402,PhysRevB.91.134423,SuzukiPhysRevB.90.184414,WhitePhysRevB.84.245130,HaukePhysRevB.87.014415}. A high magnetic field phase diagram study of the TLAF has also been performed~\cite{StarykhPhysRevLett.113.087204}. An appropriate theoretical treatment of interactions in a TLAF must consider spin wave quantum fluctuation effects~\cite{JohnPhysRevB.75.174447}. Zero-point quantum fluctuations of a noncollinear ordered quantum magnet gives rise to spin Casimir effect~\cite{PhysRevB.92.214409,PhysRevB.94.134416}. As a spin analog of the Casimir effect in vacuum, the spin Casimir effect describes the various macroscopic Casimir forces and torques that can potentially emerge from the quantum spin system. The physical consequence of the Casimir torque, generated due to the underlying lattice anisotropy, is the modification of the ordering wave vector, which is much smaller than the classical value. The modification in the ordering wave vector can cause the spin spiral state to become unstable, in turn rendering the standard spin wave theory expansion (1/S-SWT) approach inapplicable. Thus, the generic interacting spin wave theory is not appropriate.

To remedy the effect of singular behavior (which is not a precursor to the onset of quantum disordered phases) that naturally arises in noncollinear systems due to the presence of spin Casimir torque, Du \emph{et.~al.}~\cite{PhysRevB.92.214409,PhysRevB.94.134416}, proposed the torque equilibrium spin-wave theory (TESWT). The regularization scheme of TESWT formalism removes the naturally occuring divergences within the interacting 1/S-SWT formalism of the anisotropic quantum lattice model. It was shown that TESWT gives a much closer final ordering vector to the results of series expansion (SE) and modified spin wave theory (MSWT) method~\cite{hauke2011modified,PhysRevB.59.14367}. Furthermore, its prediction of the phase diagram is consistent with the previous numerical studies~\cite{hauke2011modified,PhysRevB.59.14367}.

Historically, the concept of a roton minimum and a rotonlike point in the TLAF was introduced by Zheng \emph{et.~al.}~\cite{PhysRevLett.96.057201,ZhengPhysRevB.74.224420}. Using SE method the authors identified a local minimum in the magnon dispersion at the high symmetry $M^\prime$ point, ($\pi$, $\pi/\sqrt{3}$). Drawing analogy with the appearance of a similar dip (local minimum) that is observed in the excitation spectra of superfluid $^4$He~\cite{book:780611} and the fractional quantum Hall effect~\cite{PhysRevB.33.2481}, the authors proposed the ``roton" nomenclature to describe the minimum in the magnon dispersion. The dip in the spectrum is also present at the other high symmetry $M$ point, $(0,2\pi/\sqrt{3})$, in the middle of the Brillouin zone (BZ) face edge. Zheng \emph{et.~al.} noted that a roton minimum is absent in the linear spin wave theory (LSWT) spectrum. Thus, the occurence of the rotonlike point is a consequence of quantum fluctuations arising in a frustated magnetic material~\cite{KuboPhysRevB.90.014421,PowaskiPhysRevLett.115.207202}. In a subsequent publication the concept of the rotonlike point was extended to the case of an anisotropic lattice by Fjaerestad \emph{et.al.}~\cite{JohnPhysRevB.75.174447}. Additionally, a square lattice system with $J^{\prime}/J > 2$ has also been predicted to support the roton minima~\cite{PhysRevLett.96.057201,KuboPhysRevB.90.014421}.

Further support of the roton feature was provided by the 1/S-SWT study of Starykh \emph{et.al.} \cite{StarykhPhysRevB.74.180403}. Based on their work it was proposed that rotons are part of a global renormalization (weak local minimum), with large regions of (almost) flat dispersion. The appearance of rotonlike minima and what was dubbed as a roton excitation has also been studied in an anisotropic spin-1/2 TLAF from the perspective of an algebraic vortex liquid theory \cite{AliceaPhysRevB.73.174430,AliceaPhysRevB.75.144411}. Several anomalous roton minima were predicted in the excitation spectrum in the regime of lattice anisotropy where the canted Neel state appears. From the perspective of the algebraic vortex liquid theory formulated in terms of fermionic vortices in a dual field theory, it was proposed that the roton is a vortex anti-vortex excitation, thereby, lending credence to use of the word roton as an apt description. Rotons have also been predicted to exist in field induced TLAF magnetic systems \cite{MaksimovPhysRevB.94.140407}. The field-induced transformations in the dynamical response of the XXZ model create the appearance of rotonlike minima at the $K$ point. Experimental evidence of the rotonlike point can be found in recent inelastic neuron scattering (INS) spectrum of the $\alpha$-CaCr$_{2}$O$_{2}$ system \cite{TothPhysRevB.84.054452,TothPhysRevLett.109.127203}. Examples of TLAF where anisotropy and DM interaction are present are plethora~\cite{PhysRevB.77.174412,PhysRevB.87.174423,JohnPhysRevB.75.174447,PhysRevLett.108.057205,PhysRevB.67.104431,KadowakiJPSJ,IshiiEPL,PhysRevB.81.104411,TothPhysRevB.84.054452,TothPhysRevLett.109.127203,PhysRevLett.86.1335}. %We also utilize parameters from Cs$_2$CuCl$_4$ to highlight the effects of DM interaction on the spectrum even though Cs$_2$CuCl$_4$ is on the cusp of a set of loosely coupled 1d chains~\cite{PhysRevB.74.180403}. We show through our calculations interesting physical trends which can be discerned from the bimagnon $K$ -edge indirect RIXS spectra \cite{RevModPhys.83.705} that point to highly non-trivial nature. And the new messages from experiment may challenge the modified $t$-$J$ model and the concept that overdoped cuprates are close to the regime of normal Fermi liquid.

With advancements in instrumental resolution of the next-generation synchrotron radiation sources, RIXS spectroscopy presents itself as a novel experimental tool to investigate the nature of the bimagnon RIXS spectrum and the influence of the roton~\cite{DEAN20153}. As a spectroscopic technique RIXS has the ability to probe both single-magnon and multimagnon excitations across the entire BZ \cite{RevModPhys.83.705,PhysRevB.75.214414,PhysRevLett.105.167404}. Using RIXS it is possible to probe high energy excitations in cuprates~\cite{ChenPhysRevB.88.184501,PhysRevB.86.125103}. Considering the physical behavior that has been studied within the context of RIXS TLAF and the fact that departures from the isotropic triangular lattice geometry is a norm in a frustrated TLAF, this begs the question $-$ ``What is the influence of spatial anisotropy and DM interaction on the bi- and trimagnon $K$ -edge indirect RIXS bimagnon spectrum at the rotonlike points and the other BZ points of an anisotropic triangular lattice ?"
%%%%%%%%%%%%%%%%%%%%%%%%%%%%%%%%
\begin{figure}
\centering\includegraphics[scale=0.3]{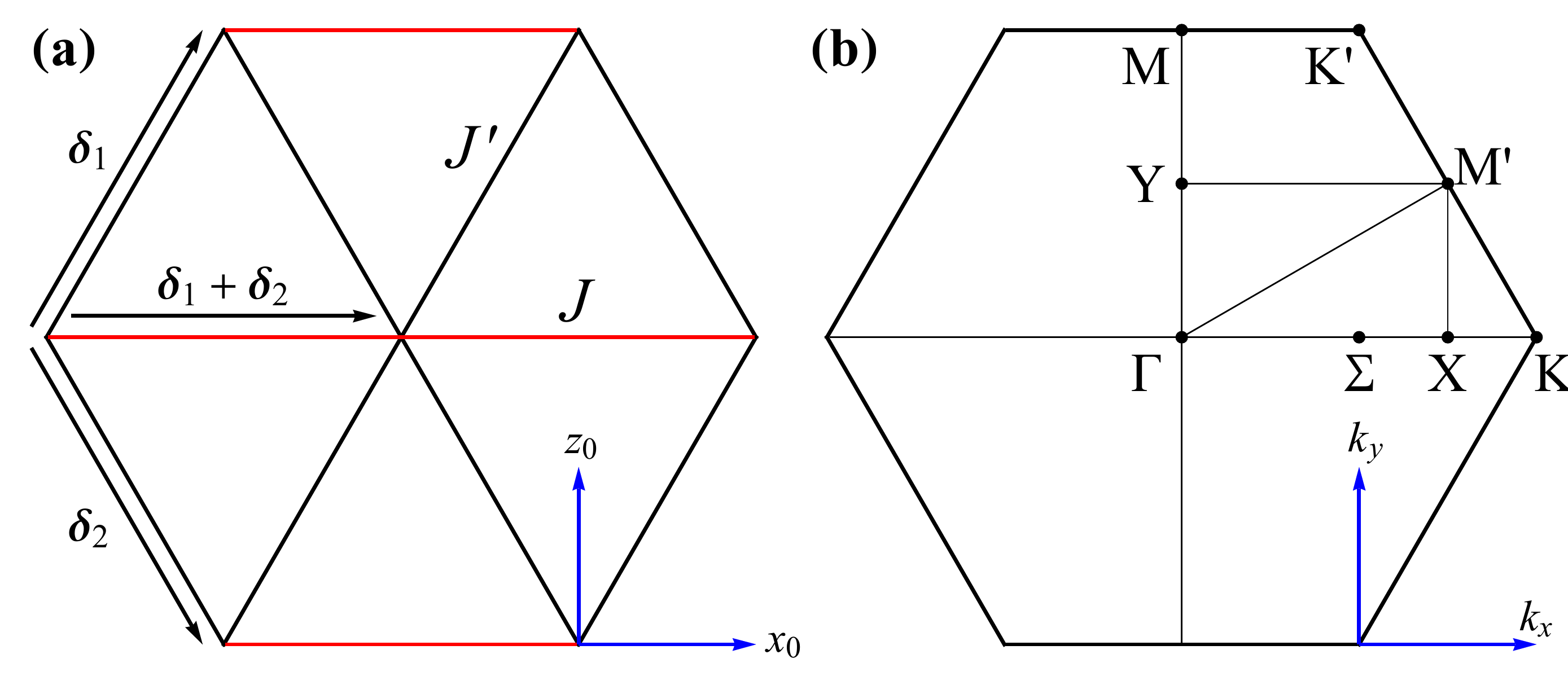}
\caption{Sketch of the triangular lattice and the Brillouin zone. (a) The anisotropic triangular lattice with exchange constant $J$ along the horizontal bonds and $J^\prime$ along the diagonal (zigzag) bonds. The lattice vectors are denoted by $\bdelta_{1,2}$. (b) The first Brillouin zone and the high-symmetry points defined as $\Gamma = (0,0)$, $\Sigma = (2\pi/3,0)$, $X = (\pi,0)$, $K = (4\pi/3,0)$, $K^\prime = (2\pi/3,2\pi/\sqrt 3)$, $M^\prime = (\pi,\pi/\sqrt 3)$, $M = (0,2\pi/\sqrt 3)$ and $Y = (0,\pi/\sqrt 3)$. The choice of co-ordinate orientation is in keeping with the convention adopted in Ref.~\cite{PhysRevB.79.144416,PhysRevB.92.035109}.
}
\label{Fig1}
\end{figure}
%%%%%%%%%%%%%%%%%%%%%%%%%%%%%%%%%%%

In this article, we utilize material parameters relevant to Cs$_2$CuCl$_4$ to elucidate the $K$ -edge RIXS behavior of the rotonlike points and also the bimagnon behavior at the $Y$ point. We apply TESWT to our quantum Heisenberg model with spatial anisotropy and DM interaction on a triangular lattice. Using a TESWT upto first order in $1/S$, we compute the final ordering vector, the spin-wave energy, and phase diagram with different anisotropy parameters. We find the phase diagram has a physically consistent behavior in the ordering wave vector $\bq$. We find that the presence of a relatively small DM interaction can make the spiral state more stable. We calculate the interplay of x-ray scattering and bi- and trimagnon excitation. We find that the evolution of the RIXS spectra at rotonlike points is non-trivial. In the isotropic case all the rotonlike points are identical due to the 60$^{\circ}$ rotation symmetry of the underlying isotropic triangular lattice. However, in the presence of symmetry breaking DM interaction terms the equivalence breaks down to give rise to two distinct points $-$ $M$ and $M^\prime$, see Fig.~\ref{Fig9}. Thus we investigate and track the evolution of the spectra at these two points separately. With increasing anisotropy the spectral weight at these points are subdued, even though the rotonlike points lie outside the region of magnon damping. Additionally, we find that the RIXS spectrum at the rotonlike $M$ point undergoes a spectral downshift. However, for the $M^\prime$ point the location of the peak is stable, albeit suppressed as the strength of the perturbation is increased. We also track the bimagnon RIXS evolution at the $Y$ point in the Brillouin zone to compare and contrast with the behavior at the rotonlike points. The spectrum at Y shows more peaks than at $M$ or $M^\prime$. Thus, the roton excitation spectrum is more stable~\cite{PhysRevB.92.035109}.

This paper is organized as follows. In Sec.~\ref{sec:model} we introduce the model spin-1/2 anisotropic TLAF with DM interaction. In  Sec.~\ref{subsec:spinexp} we state the spin wave formalism required to compute the wave vector renormalization (Sec.~\ref{subsec:renorm}) and renormalized dispersion (Sec.~\ref{subsec:cubicquartic}). In Sec.~\ref{sec:TESWT}, we extend the applicability of the TESWT formalism to include the effects of DM interaction. In Sec.~\ref{subsec:methodteswt}, we elaborate on the TESWT method, compute the ordering vector and dispersion, and perform a TESWT INS fitting (Sec.~\ref{subsec:insfit}). We then calculate the phase diagram in Sec.~\ref{subsec:S}. In Sec.~\ref{sec:rixs} we compute the indirect RIXS spectra. In Sec.~\ref{subsec:rixsa} we compute the non-interacting bi- and trimagnon spectrum. In Sec.~\ref{subsec:rixsb} we outline the formalism to compute the interacting bimagnon RIXS spectrum by including the quartic interactions. In Sec.~\ref{subsec:rixsc} we track the evolution of the roton energy to provide a physical explanation of the trend exhibited by the RIXS spectrum with anisotropy and DM interaction. In Sec.~\ref{subsec:totrixs} we state the results for the total indirect $K$ -edge RIXS intensity. Finally, in Sec.~\ref{sec:Conclu} we provide our conclusions.

%%%%%%%%%%%%%%%%%%%%%%%%%%%%%%%%%%%%%%%%%%%%%%%%%%%%%%%%%%%%%%%%
\section{Model Hamiltonian}\label{sec:model}
%%%%%%%%%%%%%%%%%%%%%%%%%%%%%%%%%%%%%%%%%%%%%%%%%%%%%%%%%%%%%%%
The antiferromagnetic Heisenberg model on the anisotropic triangular-lattice is widely believed to be well described by Cs$_2$CuBr$_4$ and Cs$_2$CuCl$_4$~\cite{PhysRevB.93.085111}. While Cs$_2$CuCl$_4$ exhibits spin-liquid behavior over a broad temperature range~\cite{PhysRevLett.86.1335,PhysRevB.68.134424}, the Cs$_2$CuBr$_4$ compound exhibits a magnetically ordered ground state with spiral order in zero magnetic field~\cite{PhysRevB.67.104431}. For $\alpha$$-$CaCr$_{2}$O$_{4}$, though it is reported to have two inequivalent Cr$^{3+}$ ions and four different exchange interactions, the nature of the distortion is such
that the average of the exchange interactions along any direction is approximtely equal.

We consider the spin-1/2 antiferromagnetic Heisenberg model on the anisotropic triangular-lattice perturbed by a DM interaction, described by the Hamiltonian
\begin{equation}\label{model}
  \mathcal{H}=\sum_{\langle ij\rangle}J_{ij}\bs_i\cdot\bs_j+H_{\mathrm{DM}},
\end{equation}
where $\langle ij\rangle$ refers to nearest-neighbor bonds on the triangular lattice, $J_{ij}=J$ denotes the exchange constants along the horizontal bonds and
$J_{ij}=J^\prime$ the diagonal bonds, see Fig.~\ref{Fig1}. The asymmetric DM interaction between neighboring spins is given by
\begin{equation}\label{dmterm}
  H_{\mathrm{DM}}=-\sum_i\mathbf{D}\cdot[\bs_i\times(\bs_{i+\bdelta_1}+\bs_{i+\bdelta_2})],
\end{equation}
where $\mathbf{D}=(0,D,0)$ with $(D>0)$ and $\bdelta_{1,2}$ are the nearest neighbor vectors along the diagonal bonds as shown in Fig.~\ref{Fig1}. In the classical limit, the spin operators are replaced by the three-component vectors
\begin{equation}
  \bs_i/S=\cos(\bq\cdot\mathbf{r}_i)\hat{z}_0+\sin(\bq\cdot\mathbf{r}_i)\hat{x}_0,
\end{equation}
where the spin forms a spiral with the ordering vector $\bq$.
The classical ground state energy is given by
\begin{gather}
  E_0(\bq)=3NJS^2(\lambda_\bq-\eta_\bq)=3NJS^2\gamma_\bq,
\end{gather}
with
\begin{gather}
  \lambda_\kk=\frac{1}{3}(\cos k_x+2\alpha\cos\frac{k_x}{2}\cos\frac{\sqrt{3}}{2}k_y),\\
  \eta_\kk=\frac{2}{3}\eta\sin\frac{k_x}{2}\cos\frac{\sqrt{3}}{2}k_y,
\end{gather}
where the dimensionless ratios $\alpha=J^\prime/J$ and $\eta=D/J$ denote the relative interaction strengths.
For the determination of the ordering vector $\bq$ we have to minimize the classical ground state energy
\begin{equation}
   \nabla_\bq E_0(\bq)=0,
\end{equation}
which amounts to finding the roots of the equations
\begin{equation}\label{Eq.8}
\left\{
 \begin{array}{cc}
  \sin Q_x+\alpha\sin\frac{Q_x}{2}\cos\frac{\sqrt{3}}{2}Q_y+\eta\cos\frac{Q_x}{2}\cos\frac{\sqrt{3}}{2}Q_y=0,\\
  \alpha\cos\frac{Q_x}{2}\sin\frac{\sqrt{3}}{2}Q_y-\eta\sin\frac{Q_x}{2}\sin\frac{\sqrt{3}}{2}Q_y=0.
   \end{array}\right.
\end{equation}
Anticipating that this condition leads to a spiral along the $x$ axis $\bq=(Q_0,0)$, we obtain the solution in the absence of DM interaction as
\begin{equation}
 Q_0=\left\{
 \begin{array}{cc}
   2\arccos(-\frac{\alpha}{2}), & \alpha<2, \\
   2\pi, & \alpha\geq 2.
 \end{array}\right.
\end{equation}
Apriori, it is not clear whether the classical ordering vector correctly describes the long-ranger order in the quantum frustrated system.
In fact, the classical wave vector will be renormalized by quantum fluctuations as will be discussed in Sec. \ref{subsec:methodteswt}.
%%%%%%%%%%%%%%%%%%%%%%%%%%%%%%%%%%%%%%%%%%%%%%%%%%%%%%%%%%%%%%%%
\section{Linear Spin-wave theory}\label{sec:spin-wave}
%%%%%%%%%%%%%%%%%%%%%%%%%%%%%%%%%%%%%%%%%%%%%%%%%%%%%%%%%%%%%%%%

\subsection{$1/S$ expansion}\label{subsec:spinexp}
Before we set up the spin-wave expansion, it is convenient to transform the spin components from the laboratory frame $(x_0,z_0)$ to the rotating frame $(x,z)$ through
\begin{gather}
  S_i^{x_0}=S_i^z\sin\theta_i+S_i^x\cos\theta_i,\\
  S_i^{z_0}=S_i^z\cos\theta_i-S_i^x\sin\theta_i,
\end{gather}
where $\theta_i=\bq\cdot\mathbf{r}_i$. The rotating Hamiltonian takes the form
\begin{gather}
 \mathcal{H}=\sum_{\langle ij\rangle}\Big[J_{ij}S_i^yS_j^y+J_{ij}^+(S_i^zS_j^z+S_i^xS_j^x)\non\\ \label{localham}
 +J_{ij}^-(S_i^zS_j^x-S_i^xS_j^z)\Big],
\end{gather}
where we have defined
\begin{align}
    J_{ij}^+=&J_{ij}\cos(\theta_i-\theta_j)+D_{ij}\sin(\theta_i-\theta_j),\\
    J_{ij}^-=&J_{ij}\sin(\theta_i-\theta_j)-D_{ij}\cos(\theta_i-\theta_j).
\end{align}

SWT amounts to applying the Holstein-Primakoff (HP) transformation to bosonize the rotating Hamiltonian (\ref{localham})
\begin{equation}
 S_i^z=S-n_i,\ S_i^-=a^\dag\sqrt{2S-n_i},\ S_i^+=(S_i^-)^\dag,
\end{equation}
where $n_i=a_i^\dag a_i$ and $a_i^\dag$ ($a_i$) is the magnon creation (annihilation) operator for a given site $i$.
Under the assumption of diluteness of the HP boson gas, $n_i/(2S) \ll 1$,
one arrives at the interacting spin-wave Hamiltonian to the first order expansion of the square root
\begin{equation}\label{swham}
 \mathcal{H}=E_0(\bq)+H_2+H_3+H_4,
\end{equation}
where the first term is the classical energy and $H_n$ denotes terms of the $n^{th}$ power in the HP boson operators $a^\dag(a)$.

\subsection{Quadratic terms: first-order corrected LSWT}\label{subsec:renorm}
After Fourier transformation we obtain the quadratic Hamiltonian in momentum space as
\begin{equation}
    H_2=\sum_\kk \Big[A_\kk a_\kk^\dag a_\kk+\frac{B_\kk}{2}(a_\kk^\dag a_{-\kk}^\dag+a_{-\kk}a_\kk)\Big],
\end{equation}
with
\begin{gather}
  A_\kk=3JS[\lambda_\kk+\xi_\kk-2\gamma_\bq],\non\\
  B_\kk=3JS[\xi_\kk-\lambda_\kk],
\end{gather}
where
\begin{align}
    \xi_\kk=&\frac{1}{2}(\gamma_{\bq+\kk}+\gamma_{\bq-\kk}).
\end{align}
Diagonalization of $H_2$ is performed with the canonical Bogoliubov transformation
\begin{equation}
 a_\kk=u_\kk b_\kk+v_\kk b_{-\kk}^\dag,
\end{equation}
with the parameters $u_{\kk}$ and $v_{\kk}$ defined as
\begin{equation}
u_{\kk}=\sqrt{\frac{A_\kk+\varepsilon_{\kk}}{2\varepsilon_{\kk}}},\
v_{\kk}=-\frac{B_\kk}{|B_\kk|}\sqrt{\frac{A_\kk-\varepsilon_{\kk}}{2\varepsilon_{\kk}}}.
\end{equation}
As a result we obtain the linear spin-wave dispersion
\begin{equation}
 \varepsilon_\kk=\sqrt{A_\kk^2-B_\kk^2}.
\end{equation}
It is noted that the magnon spectrum has zeros at $\kk=0$ while a gap is opened at $\kk=\bq$ in the presence of DM interaction.
The diagonalized Hamiltonian $H_2$ is given by
\begin{equation}
    H_2=E_2(\bq)+\sum_\kk \varepsilon_\kk b_\kk^\dag b_\kk,
\end{equation}
where the zero-point energy
\begin{equation}
    E_2(\bq)=3NSJ\gamma_\bq+\frac{1}{2}\sum_\kk \varepsilon_\kk,
\end{equation}
is the $1/S$ correction to the classical ground-state energy.
%%%%%%%%%%%%%%%%%%%%%%%%%%%%%%%%%%%%%%%%%%%
\begin{figure}
\centering\includegraphics[width=3.4in]{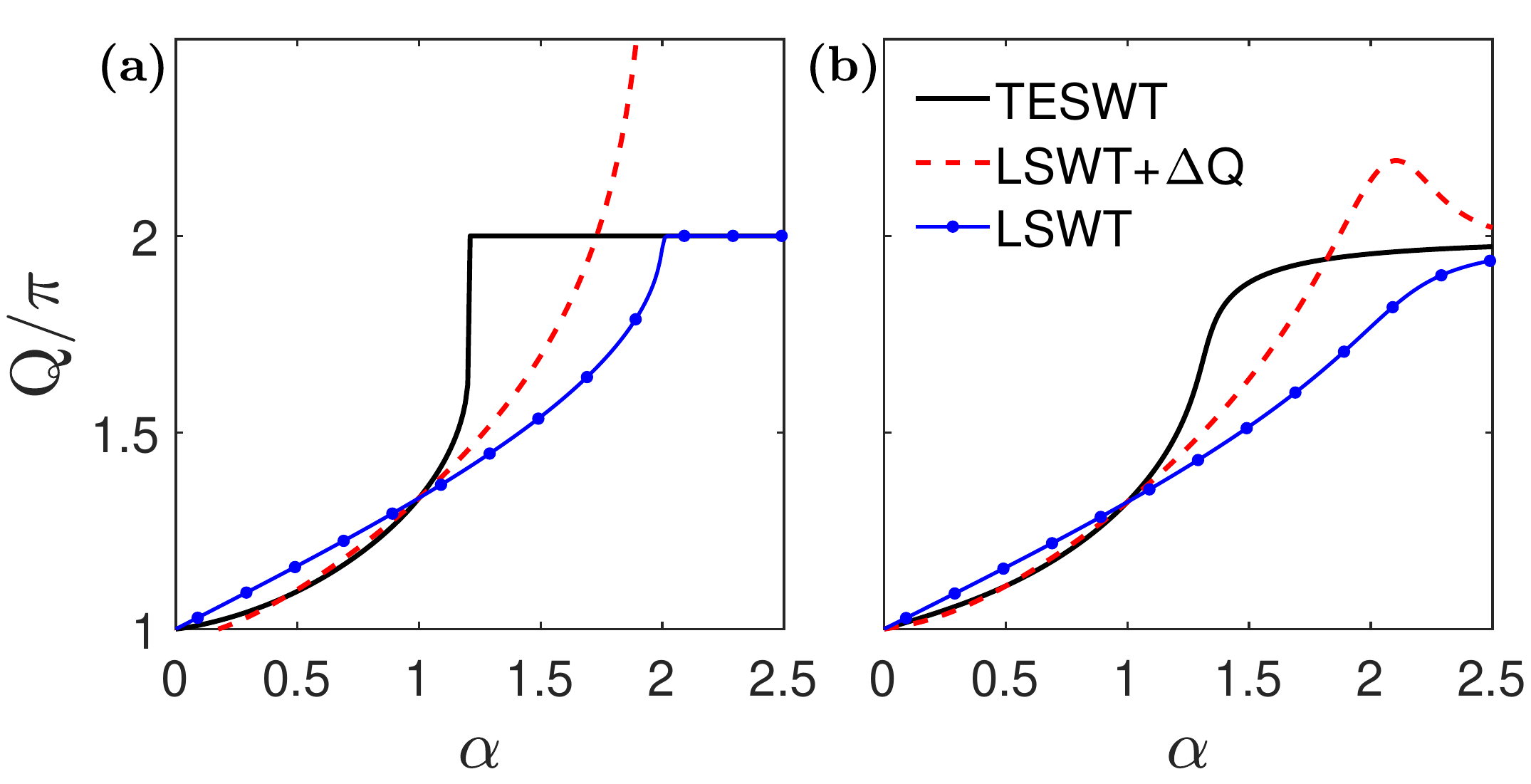}
\caption{The evolution of ordering wave vector $Q$ for the $S=\frac{1}{2}$ spiral antiferromagnet on the anisotropic triangular lattice as a function of $\alpha=J^{\prime}/J$. The ordering vectors of TESWT, LSWT, and $1/S$ corrected LSWT are compared. (a) $\eta=D/J=0$, (b) $\eta=D/J=0.05$.}
\label{Fig2}
\end{figure}
%%%%%%%%%%%%%%%%%%%%%%%%%%%%%%%%%%%%%%%%%%
Generally, the first-order correction of LSWT $\bq_1=\bq_0+\Delta\bq$ is determined by minimizing the sum $E_0(\bq)+E_2(\bq)$
\begin{equation}
   \nabla_\bq[E_0(\bq)+E_2(\bq)]=0.
\end{equation}
Neglecting higher order terms, we obtain
\begin{align}
\nabla_\bq[E_0(\bq_1)+E_2(\bq_1)]=\nabla_\bq E_2(\bq_0)+\Delta\bq\cdot K=0,
\end{align}
with
\begin{align}
  K_{\alpha,\beta}=\frac{\partial^2E_0(\bq_0)}{\partial Q_\beta Q_\alpha}.
\end{align}
A straightforward calculation gives $1/S$ correction to the classical wave vector
\begin{equation}
    \Delta \boldsymbol{Q}=-\boldsymbol{w}\cdot K^{-1},
\end{equation}
where
\begin{equation}
    w_{\alpha}=\frac{\partial E_2(\bq_0)}{\partial Q_\alpha}.
\end{equation}
In Fig.~\ref{Fig2} we display the variation of the ordering wave vector renormalization against  lattice anisotropy computed using LSWT, 1/S corrected LSWT, and TESWT. It is clear that while the LSWT formulation extends the spiral phase region, the first-order correction from 1/S-LSWT gives an unphysical result as $\alpha \rightarrow 2$ while $\eta=0$. Inclusion of DM interaction rounds the singularity with an angle that is greater than
$2\pi$. The root cause of this divergence originates from spin Casimir torque \cite{PhysRevB.92.214409,PhysRevB.94.134416}. In a frustrated spiral system, the strong quantum fluctuation effect leads to failure in the first-order correction. In Sec.~\ref{sec:TESWT} we will discuss and implement the TESWT approach which offers a solution to this issue. The equations to generate the TESWT results are reported in that section.
%We display the $1/S$ corrected LSWT ordering wave vector renormalization in Fig.~\ref{Fig2}. It is clear that such first-order correction gives an unphysical result as it become infinity as $\alpha\to2$ while $\eta=0$. In such a frustrated system, the strong quantum fluctuation effect lead the first-order
%correction to failure. We will discuss this in Sec.~\ref{sec:TESWT}.

\subsection{Cubic and quartic terms: renormalized dispersion}\label{subsec:cubicquartic}
The 1/S correction to the spin wave dispersion has to be accounted for in a non-collinear structure. The
interplay of magnon decay as it arises from the non-collinear structure is also considered \cite{RevModPhys.85.219,PhysRevLett.97.207202,PhysRevB.88.094407}.
The three-boson term that arises from the coupling between transverse and longitudinal fluctuations in the noncollinear spin structure takes the form~\cite{PhysRevB.79.144416},
\begin{align}
    H_3=-\sqrt{\frac{S}{2}}\sum_{\langle ij\rangle}&J_{ij}^-[a^\dag_ia_i(a^\dag_j+a_j)-a^\dag_ja_j(a^\dag_i+a_i)].
\end{align}
In momentum space, we obtain
\begin{equation}
    H_3=\frac{3JSi}{2}\sqrt{\frac{3}{2SN}}\sum_{1+2=3}(\bar{\gamma}_1+\bar{\gamma}_2)(a^\dag_1a^\dag_2a_3-a^\dag_3a_1a_2),
\end{equation}
where we have defined
\begin{equation}
  \bar{\gamma}_\kk=\frac{1}{\sqrt{3}}(\gamma_{\bq+\kk}-\gamma_{\bq-\kk}).
\end{equation}
In the above we have adopted the convention that $1=\kk_1$, $2=\kk_2$, etc. For example, $a_1\equiv a_{\kk_1}$.
Performing the Bogoliubov transformation in $H_3$ we obtain the interaction terms expressed via the magnon operators as
\begin{eqnarray}
   H_3=&&\frac{1}{2!}\sum_{1+2=3}V_{a}(1,2;3)(b_{1}^\dag b_{2}^\dag b_{3}+\mathrm{H.c.})\non\\
 &&+\frac{1}{3!}\sum_{1+2+3=0}V_{b}(1,2,3)(b_{1}^\dag b_{2}^\dag b_{3}^\dag+\mathrm{H.c.}).
\end{eqnarray}
The three-boson vertices are given by
\begin{equation}
  V_{a,b}(1,2;3)=3Ji\sqrt{\frac{3S}{2N}}\bar{V}_{a,b}(1,2;3),
\end{equation}
with $\bar{V}_{a,b}$ given by
\begin{eqnarray}
  \bar{V}_{a}(1,2;3)=&&\bar{\gamma}_1(u_1+v_1)(u_2u_3+v_2v_3)+\bar{\gamma}_2(u_2+v_2)(u_1u_3\non\\
  &&+v_1v_3)-\bar{\gamma}_3(u_3+v_3)(u_1v_2+v_1u_2),\\
  \bar{V}_{b}(1,2,3)=&&\bar{\gamma}_1(u_1+v_1)(u_2v_3+v_2u_3)+\bar{\gamma}_2(u_2+v_2)(u_1v_3\non\\
  &&+v_1u_3)+\bar{\gamma}_3(u_3+v_3)(u_1v_2+v_1u_2).
\end{eqnarray}
We notice that the three-magnon vertices are of order $1/\sqrt{S}$ relative to the linear spin-wave Hamiltonian and they must occur in pairs in any
self-energy or polarization diagram.
The quartic term $H_4$ in the interacting spin-wave Hamiltonian (\ref{swham}) reads
\begin{eqnarray}
  H_{4}=\sum_{\langle ij\rangle}&&\Big[\frac{1}{2}J_{ij}^+a^\dag_ia_ia^\dag_ja_j+\frac{1}{8}(J_{ij}-J_{ij}^+)(a^\dag_ia_ia_ia_j+a^\dag_ja_ja_ja_i)\non\\
                 &&-\frac{1}{8}(J_{ij}+J_{ij}^+)(a^\dag_ja^\dag_ia_ia_i+a^\dag_ja^\dag_ja_ja_i)\Big]+\mathrm{H.c.}
\end{eqnarray}
To obtain the explicit forms of the quasiparticle representation of $H_4$, we introduce the following mean-field averages
\begin{gather}
  n_\kk=\langle a_\kk^\dag a_\kk \rangle=\frac{A_\kk-\varepsilon_\kk}{2\varepsilon_\kk},
  \Delta_\kk=\langle a_\kk a_{-\kk} \rangle=-\frac{B_\kk}{2\varepsilon_\kk}.
\end{gather}
The Hartree-Fock decoupling of the $H_4$ yields the quadratic Hamiltonian
\begin{eqnarray}
  \delta H_2=\sum_\kk \Big[\delta A_\kk a_\kk^\dag a_\kk+\frac{1}{2}\delta B_\kk(a_\kk^\dag a_{-\kk}^\dag+a_{-\kk}a_\kk)\Big],
\end{eqnarray}
where
\begin{eqnarray}
  \delta A_\kk=&&A_\kk+\frac{1}{2SN}\sum_\qq \frac{1}{\varepsilon_\qq}\bigg[A_\qq \Big(A_{\kk-\qq}+B_{\kk-\qq}-A_\kk-A_\qq \Big)\non\\
  &&+B_\qq \Big(\frac{B_\kk}{2}+B_\qq \Big) \bigg],\\
  \delta B_\kk=&&B_\kk-\frac{1}{2SN}\sum_\qq \frac{1}{\varepsilon_\qq}\bigg[B_\qq\Big(A_{\kk-\qq}+B_{\kk-\qq}-\frac{A_\kk}{2}-\frac{A_\qq}{2}\Big)\non\\
  &&+A_\qq\Big(B_\kk+\frac{B_\qq}{2}\Big)\bigg],
\end{eqnarray}
We then obtain the Hartree-Fock corrected $H_2$ term as
\begin{eqnarray}
  \delta H_2=\sum_\kk \Big[\delta\varepsilon_\kk b_\kk^\dag b_\kk+\frac{O_\kk}{2}(b_\kk^\dag b_{-\kk}^\dag+b_\kk b_{-\kk})\Big],
\end{eqnarray}
where
\begin{eqnarray}
 \delta\varepsilon_\kk=(u_\kk^2+v_\kk^2)\delta A_\kk+2u_\kk v_\kk\delta B_\kk,\\
 O_\kk=(u_\kk^2+v_\kk^2)\delta B_\kk+2u_\kk v_\kk\delta A_\kk.
\end{eqnarray}
Finally, the normal-ordered quartic term $\tilde{H}_4$ in the quasiparticle representation describes the multi-magnon interactions.
In the hierarchy of $1/S$ expansion, terms relevant for our calculations are the lowest order irreducible two-magnon scattering amplitude
\begin{eqnarray}
     \tilde{H}_4^{2-p}=\sum_{\kk_1+\kk_2=\kk_3+\kk_4}V_{c}(\kk_1,\kk_2;\kk_3,\kk_4)b_{\kk_1}^\dag b_{\kk_2}^\dag b_{\kk_3}b_{\kk_4},
\end{eqnarray}
with the vertex function given by
\begin{widetext}
\begin{eqnarray}
  V_{c}(1,2;3,4)=&&\frac{1}{8SN}\Big\{-(B_1+B_2+B_4)(u_1u_2u_3v_4+v_1v_2v_3u_4)
                 -(B_1+B_2+B_3)(u_1u_2v_3u_4+v_1v_2u_3v_4)\non\\
                 &&-(B_2+B_3+B_4)(u_1v_2u_3u_4+v_1u_2v_3v_4)
                 -(B_1+B_3+B_4)(u_1v_2v_3v_4+v_1u_2u_3u_4)\non\\
                 &&+[(C_{1-3}+C_{2-3}+C_{1-4}+C_{2-4})
                 -(A_1+A_2+A_3+A_4)](u_1u_2u_3u_4+v_1v_2v_3v_4)\non\\
                 &&+[(C_{1+2}+C_{3+4}+C_{1-3}+C_{2-4})
                 -(A_1+A_2+A_3+A_4)](u_1v_2u_3v_4+v_1u_2v_3u_4)\non\\
                 &&+[(C_{1+2}+C_{3+4}+C_{1-4}+C_{2-3})
                 -(A_1+A_2+A_3+A_4)](u_1v_2v_3u_4+v_1u_2u_3v_4)\Big\},
\end{eqnarray}
\end{widetext}
where we have defined
\begin{eqnarray}
  C_\kk=A_\kk+B_\kk.
\end{eqnarray}

The effective $1/S$  interacting spin$-$wave Hamiltonian in terms of the magnon operators reads
\begin{eqnarray}\label{effectiveham}
 \mathcal{H}_{\mathrm{eff}}=&&\sum_\kk\Big[(\varepsilon_\kk+\delta\varepsilon_\kk)b_\kk^\dag b_\kk+\frac{O_\kk}{2}(b_\kk^\dag b_{-\kk}^\dag+b_\kk b_{-\kk})\Big]\non\\
 &&+\frac{1}{2!}\sum_{\{\kk_i\}}V_{a}(b_{1}^\dag b_{2}^\dag b_{3}+\mathrm{H.c.})+\frac{1}{3!}\sum_{\{\kk_i\}}V_{b}(b_{1}^\dag b_{2}^\dag b_{3}^\dag+\mathrm{H.c.})\non\\
 &&+\sum_{\{\kk_i\}}V_{c}b_{1}^\dag b_{2}^\dag b_{3}b_{4}.
\end{eqnarray}
At zero temperature the bare magnon propagator is defined as
\begin{eqnarray}\label{bareg}
\mathrm{G}_0^{-1}(\kk,\omega)=\omega-\varepsilon_\kk+i0^+.
\end{eqnarray}
The first order $1/S$ correction to the magnon energy is determined by the Dyson equation
\begin{eqnarray}
  \omega-\varepsilon_\kk-\Sigma(\kk,\omega)=0,
\end{eqnarray}
with the one-loop self-energy $\Sigma(\kk,\omega)=\Sigma_{a}(\kk,\omega)+\Sigma_{b}(\kk,\omega)+\Sigma_{c}(\kk)$,
where $\Sigma_{c}(\kk)=\delta\varepsilon_\kk$ is a frequency-independent Hartree-Fock correction,
while $\Sigma_{a,b}(\kk,\omega)$ are calculated as
\begin{eqnarray}\label{Eq.self-erengy1}
 \Sigma_{a}(\kk,\omega)&=&\frac{1}{2}\sum_\pp \frac{|V_{a}(\pp,\kk-\pp;\kk)|^2}{\omega-\varepsilon_\pp-\varepsilon_{\kk-\pp}+i0^+},\\ \label{Eq.self-erengy2}
 \Sigma_{b}(\kk,\omega)&=&-\frac{1}{2}\sum_\pp\frac{|V_{b}(\pp,-\kk-\pp,\kk)|^2}{\omega+\varepsilon_\pp+\varepsilon_{\kk+\pp}-i0^+} .
\end{eqnarray}
The on-shell solution consists of setting $\omega=\varepsilon_\kk$ in the self-energy Eqs.~\eqref{Eq.self-erengy1} and~\eqref{Eq.self-erengy2}
leads to the following expression for the $1/S$ renormalized spectrum
\begin{eqnarray}
  \omega_\kk\equiv\bar{\omega}_\kk-i\Gamma_\kk=\varepsilon_\kk+\Sigma(\kk,\varepsilon_\kk),
\end{eqnarray}
where $\bar{\omega}_\kk=\mathrm{Re}[\omega_\kk]$ is the renormalized spin$-$wave energy and $\Gamma_\kk=-\mathrm{Im}[\omega_\kk]$ represents the magnon decay rate. In Fig.~\ref{Fig3}, we plot the $1/S$ LSWT dispersion of Cs$_2$CuCl$_4$~\cite{JohnPhysRevB.75.174447}.

%%%%%%%%%%%%%%%%%%%%%%%%%%%%%%%%%%%%%%%%%%%%%%%%%%%%%%%%%%%%%%%%
\section{Torque Equilibrium Spin Wave Theory}\label{sec:TESWT}
%%%%%%%%%%%%%%%%%%%%%%%%%%%%%%%%%%%%%%%%%%%%%%%%%%%%%%%%%%%%%%%%
%In this paper, we use TESWT \cite{PhysRevB.92.214409,PhysRevB.94.134416} and generalize the formulation to include DM interactions.
Zero-point quantum fluctuation in a non-collinear ordered spin structure can lead to deviations in the measured ordering wave vector compared to the classical one. The correction emerging from the spin Casimir effect is usually neglected, but it was recently shown that this is not a bonafide assumption. In Du \emph{et.~al.}~\cite{PhysRevB.92.214409,PhysRevB.94.134416} it was clearly established that in certain situations a standard spin wave theory is no longer applicable due to the spin Casimir quantum effect, even when the system is long-range ordered. An important consequence of these quantum fluctuations is on the spiral state which can become unstable, which is different from the case of long-range-order melting. As mentioned earlier the classical signatures of these instabilities are the divergences of the ordering wave vector at the quantum critical point and the strongly singular one-loop expansions of the energy spectrum and the sublattice magnetization. In this section, we extend the applicability of the TESWT formalism to include the effects of DM interaction in an anisotropic TLAF. Using INS experimental data from Cs$_2$CuCl$_4$~\cite{PhysRevB.68.134424}, we obtain fitting parameters for the exchange constants and DM interactions utilized in subsequent indirect $K$ -edge RIXS calculations.

\subsection{TESWT formalism}\label{subsec:methodteswt}
Spin Casimir effect will change the classical ground state to a new saddle point. This new ground state can be unambiguously determined once we compute the value of $\bq$. An ordinary approach is considering the $1/S$ correction $\Delta Q$, as we show in Sec. \ref{subsec:renorm}. However, such a method gives an unphysical result, see Fig.~\ref{Fig2}. As $\alpha\rightarrow 2$, the $1/S$ correction $\Delta Q$ becomes infinites.

The basic idea of TESWT is to minimize the ground state energy. The spin Casimir torque is defined as
\begin{equation}
  \mathbf{T}_{sc}(\bq) = \sum_{\kk}{\Bigg\langle\Psi_{vac}\left| \frac{\partial H_{sw}}{\partial \bq}\right| \Psi_{vac}}\Bigg\rangle,
\end{equation}
where $|\Psi_{vac}\rangle$ represents the quasiparticle vacuum state. Then the torque equilibrium condition is
\begin{eqnarray}\label{Eq:TESWTcondi}
\begin{split}
  \mathbf{T}_{sc}(\bq)+\mathbf{T}_{cl}(\bq) &= \sum_{\kk}{\Bigg\langle\Psi_{vac}\left| \frac{\partial (H_{sw}+H_{cl})}{\partial \bq}\right| \Psi_{vac}\Bigg\rangle} =0,\\
  \mathbf{T}_{sc}(\bq_{cl}) &= \frac{3JS}{2}\sum_{\kk}{\frac{A_\kk-B_\kk}{\varepsilon_{\kk}}\frac{\partial \gamma_{\kk+\bq}}{\partial \bq}} \Bigg|_{\bq_{cl}},\\
  \mathbf{T}_{cl}(\bq) &= 3NJS^2\frac{\partial \gamma_\bq}{\partial \bq},\quad \mathbf{T}_{cl}(\bq_{cl}) = 0,
\end{split}
\end{eqnarray}
where $\bq$ is the final ordering vector, $H_{cl}=E_0(\bq)$ is the classical energy. Using the fact that the spin-wave spectrum function $\varepsilon_\kk$ is only well defined at $\bq_{cl}$, we try to find a system whose classical ordering vector is $\bq$ for convenience of calculation. Thus we shift the function depending on classical ordering vector $\bq_{cl}$ to $\bq$ by
\begin{eqnarray}
  H_2(\alpha,\eta,\bq)=\widetilde{H}_2(\widetilde{\alpha},\widetilde{\eta},\bq)+H_2^c, \\
  A_\kk=\widetilde{A}_\kk+A_\kk^c, \quad B_\kk=\widetilde{B}_\kk+B_\kk^c,
\label{Eq.shift}
\end{eqnarray}
where $\widetilde{H}_2, \widetilde{A}_\kk$ and $\widetilde{B}_\kk$ are functions of another spin system whose classical ordering vector $\widetilde{\bq}_{cl}$ equals $\bq$. The counterterm is given by $H^{c}_{2}$ whose effects are considered in the $A_{\kk}(B_{\kk})$ coefficients through $A^{c}_{\kk}(B^{c}_{\kk})$. In principle, we have many combinations of $(\widetilde{\alpha},\widetilde{\eta})$ that satisfy this condition. As $\eta/\alpha$ is small, within perturbation theory, we believe $\widetilde{\eta}=\eta$ is a reasonable choice. Thus the new parameters can be deduced by solving the following self-consistent equations
\begin{equation}
\begin{split}
 \left\{
 \begin{array}{l}
  \widetilde{\alpha}=-2\cos{\frac{Q}{2}}-\eta \cot{\frac{Q}{2}},\\
   \widetilde{\eta}=\eta.
   \end{array}\right.
\end{split}
\end{equation}
%%%%%%%%%%%%%%%%%%%%%%%%%%%%%%%%
\begin{figure}[t]
\centering\includegraphics[width=3.4in]{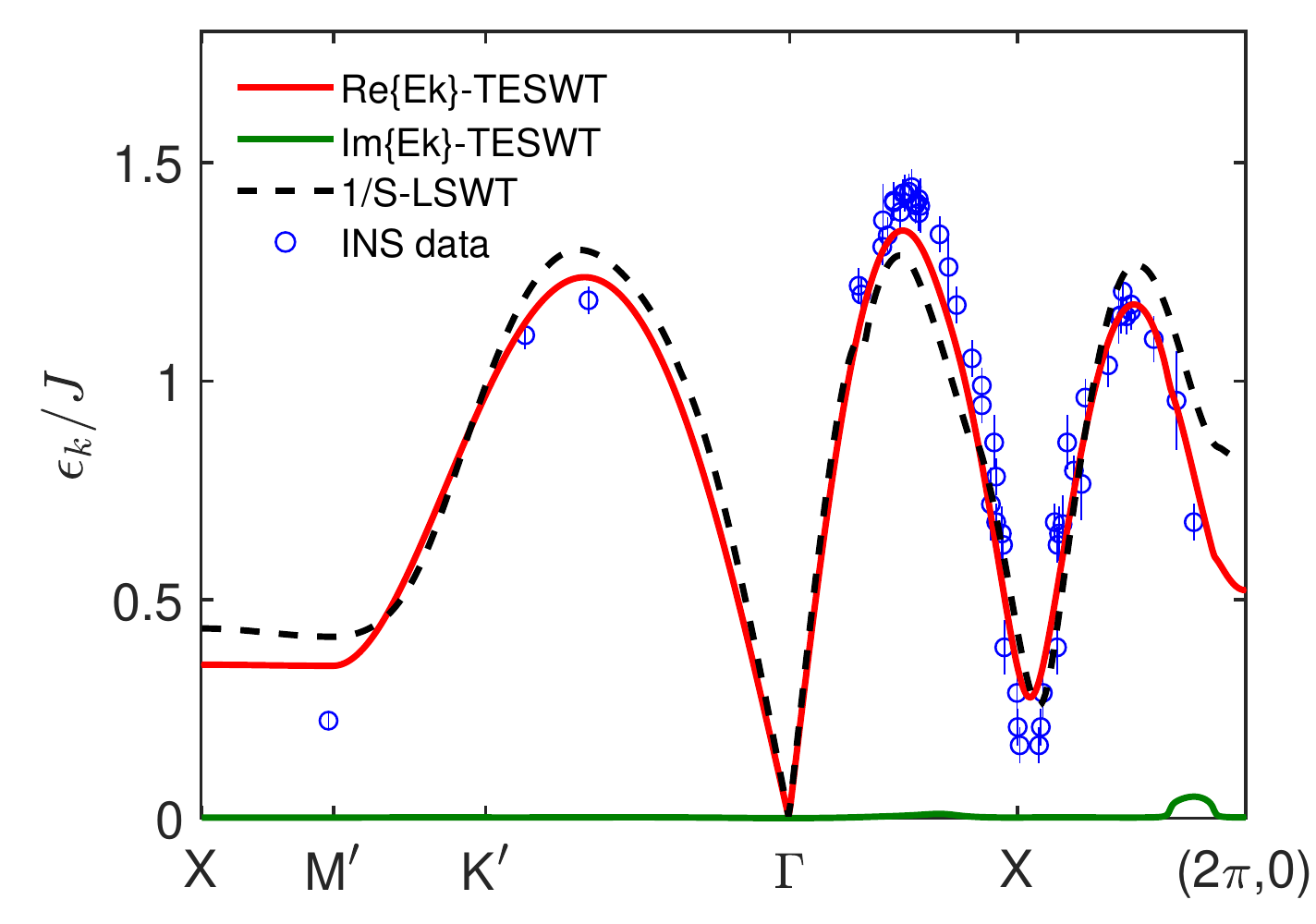}
\caption{Magnon dispersion $\epsilon_{{\bf k}}$ within TESWT and $1/S$-LSWT approach. The red line is fitted by TESWT with $\alpha=0.316$ and $\eta=0.025$ ($J=0.480(9)$meV). The circles are experimental data of inelastic neutron scattering for Cs$_2$CuCl$_4$~\cite{PhysRevB.68.134424}. The black dashed line is the fitting result of $1/S$-LSWT with $\alpha=0.417$ and $\eta=0.021$($J=0.573(9)$meV). The energies of all results are normalized by $J=0.480$ meV. The momentum points in the path are defined in Fig.~\ref{Fig1}.}
\label{Fig3}
\end{figure}
%%%%%%%%%%%%%%%%%%%%%%%%%%%%%%%%
The spin Casimir torque is then expressed approximately as $\mathbf{T}_{sc}(\bq) = \widetilde{\mathbf{T}}_{sc}(\bq)$. Thus the torque equilibrium equation in Eq.(~\ref{Eq:TESWTcondi}) can be written as
\begin{equation}
  \frac{\partial \gamma_\bq}{\partial \bq}=-\frac{1}{2NS}\sum_\kk{\frac{\widetilde{A}_\kk-\widetilde{B}_\kk}{\widetilde{\epsilon}_\kk}\cdot \frac{\partial \widetilde{\gamma}_{\kk+\bq}}{\partial \bq} }.
\end{equation}

Note, the exchange parameters on the left-hand side of the equation are exact as $\alpha,\eta$. While the parameters on the right-hand side approximate as $\widetilde{\alpha}=-2\cos{\frac{Q}{2}}-\eta \cot{\frac{Q}{2}}$. We solve this equation numerically and give the results in Fig.~\ref{Fig2}. If there is no DM interaction, TESWT gives $Q=2\pi$ for $\alpha\geq 1.2$, which are similar to the results of numerical methods~\cite{hauke2011modified,PhysRevB.59.14367}. The LSWT, however, gives a wider region for spiral order phase, can't describe the region for $1.2\leq\alpha\leq2$. As anticipated, even a small DM interaction, $\eta=0.05$, changes our final ordering vector. The DM interaction improves the spiral order stabilization and enlarges it's region of validity.

We diagonalize $\widetilde{H}_2(\widetilde{\alpha},\widetilde{\eta},\bq)$ and treat $H_2^c$ as a counterterm. Since we are considering a $1/S$ theory, we neglect the counterterm contributions from $H_3^c$ and $H_4^c$~\cite{PhysRevB.92.214409,PhysRevB.94.134416}. Thus, we can write the Hamiltonian as
\begin{equation}
  \widetilde{H}_{sw}=\widetilde{H}_2+H_2^c+\widetilde{H}_3+\widetilde{H}_4.
\label{Eq:teswtexp}
\end{equation}
%%%%%%%%%%%%%%%%%%%%%%%%%%%%%%%%%%%%%%%%%%%%%%%%%%%%%%%%%%%%
\begin{figure*}
\centering\includegraphics[width=7in]{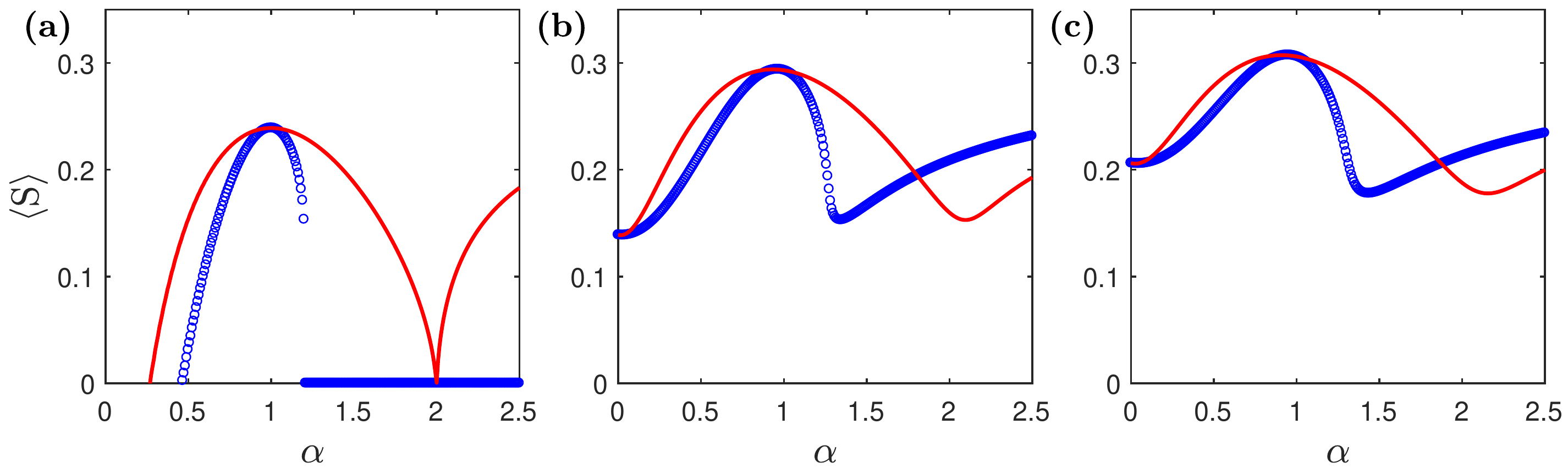}
\caption{Variation of sublattice magnetization $\langle S \rangle$ with spatial anisotropy $\alpha$. The blue (red) circles (solid line) represents TESWT (LSWT) results. (a) $\eta=0$, (b) $\eta=0.03$ and (c) $\eta=0.05$.}
\label{Fig4}
\end{figure*}
%%%%%%%%%%%%%%%%%%%%%%%%%%%%%%%%%%%%%%%%%%%%%%%%%%%%%%%%%%%%%

Following the procedure outlined in Sec.~\ref{sec:spin-wave}, the effective TESWT Hamiltonian now reads
\begin{eqnarray}\label{effectiveham}
 \widetilde{\mathcal{H}}_{\mathrm{eff}}=&&\sum_\kk\bigg[(\widetilde{\varepsilon}_\kk+\delta\widetilde{\varepsilon}_\kk)b_\kk^\dag b_\kk+\frac{\widetilde{O}_\kk}{2}(b_\kk^\dag b_{-\kk}^\dag+b_\kk b_{-\kk})\non\\
 &&+\varepsilon_\kk^c b_\kk^\dag b_\kk+\frac{O_\kk^c}{2}(b_\kk^\dag b_{-\kk}^\dag+b_\kk b_{-\kk})\bigg]\non\\
 &&+\frac{1}{2!}\sum_{\{\kk_i\}}\widetilde{V}_{a}(b_{1}^\dag b_{2}^\dag b_{3}+\mathrm{H.c.})+\frac{1}{3!}\sum_{\{\kk_i\}}\widetilde{V}_{b}(b_{1}^\dag b_{2}^\dag b_{3}^\dag+\mathrm{H.c.})\non\\
 &&+\sum_{\{\kk_i\}}\widetilde{V}_{c}b_{1}^\dag b_{2}^\dag b_{3}b_{4}.
\end{eqnarray}
where $\widetilde{F}$ means $F(\widetilde{\alpha},\widetilde{\eta},\bq)$ ($F$ is an arbitrary operator) and
\begin{gather}
\varepsilon_\kk^c=(\widetilde{u}_\kk^2+\widetilde{v}_\kk^2)A_\kk^c+2\widetilde{u}_\kk \widetilde{v}_\kk B_\kk^c
  =\frac{1}{\widetilde{\varepsilon_\kk}}\Big[\widetilde{A}_\kk A_\kk-\widetilde{B}_\kk B_\kk\Big]-\widetilde{\varepsilon}_\kk,\\
O_\kk^c=(\widetilde{u}_\kk^2+\widetilde{v}_\kk^2)B_\kk^c+2\widetilde{u}_\kk \widetilde{v}_\kk A_\kk^c
  =\frac{1}{\widetilde{\varepsilon_\kk}}\Big[\widetilde{A}_\kk B_\kk-\widetilde{B}_\kk A_\kk\Big].
\end{gather}
%%%%%%%%%%%%%%%%%%%%%%%%%%%%%%%%%%%%%%%%%%%%%%%%%%%%%
\begin{table}[t]
\caption{\label{tab:table1}
Parameter values of Cs$_2$CuCl$_4$ using different methods. The first line is our TESWT fitting results. The second line is our 1/S-SWT fitting parameters. The third line gives the fitting parameters of series expansion (SE) method~\cite{JohnPhysRevB.75.174447}. The last line gives the parameters measured by Electron-Spin-Resonance (ESR)~\cite{PhysRevLett.112.077206}.
}
\begin{ruledtabular}
\begin{tabular}{cccc}
Method & $J$(meV) & $J^\prime$(meV) & $D$(meV) \\
\colrule
TESWT & $0.480\pm0.009$ & $0.152\pm0.015$ & $0.012\pm0.002$ \\
$1/S$-LSWT & $0.573\pm0.009$ & $0.239\pm0.014$ & $0.012\pm0.001$ \\
SE & $0.374\pm0.005$ & $0.128\pm0.005$ & $0.020\pm 0.002$ \\
ESR & $0.41\pm0.02$ & $0.122\pm0.006$ & $-$ \\
\end{tabular}
\end{ruledtabular}
\end{table}
%%%%%%%%%%%%%%%%%%%%%%%%%%%%%%%%%%%%%%%%%%%%%%%%%Everything  easy to repeat as we need only to consider $H_2^c$ and $\bq$ from TESWT.
Thus, we shifted the classical ordering vector $\bq_{cl}$ to the final ordering vector $\bq$ using TESWT. Therefore, the first order $1/S$ corrected magnon dispersion can now be changed to
\begin{equation}
  \omega_\kk=\widetilde{\varepsilon}_\kk+\varepsilon_\kk^c+\delta\widetilde{\varepsilon}_\kk+\widetilde{\Sigma}_3^a(\kk,\widetilde{\varepsilon}_\kk)+\widetilde{\Sigma}_3^b(\kk,\widetilde{\varepsilon}_\kk),
\label{Eq.teswtenergy}
\end{equation}

\subsection{INS fitting}\label{subsec:insfit}
As discussed above, with anisotropy the application of 1/S-LSWT formalism is tricky. But, application of TESWT requires magnetic interaction parameters computed within that formalism. The most direct way to achieve this goal is to compare the theoretical dispersion with the experimental data. We fit the INS data of Cs$_2$CuCl$_4$~\cite{PhysRevB.68.134424} to Eq.~\eqref{Eq.teswtenergy} using iterative least squares estimation both by TESWT and $1/S$-LSWT. Our fitting parameters along with results from other sources are reported in Table.~\ref{tab:table1}. Our dispersion line fits are reported in Fig.~\ref{Fig3}. The absence of higher order terms within our TESWT could be a source of disagreement with the series expansion results~\cite{JohnPhysRevB.75.174447}, which is an all numerical method that considers higher order terms~\cite{ZhengPhysRevB.74.224420}. As the fitted dispersion by TESWT gives a reasonable comparison with the experimentally fitted SE method parameters, we believe that our TESWT can capture the essential physical behavior. While it maybe fruitful to investigate the above mentioned discrepancy, within the context of our RIXS calculation we do not expect the improved interaction constants to bring about much qualitative or quantitative differences. %Additionally, TESWT is easier to carry out and and gives the final ordering vector quickly.
%The most direct inspection method is to compare the theoretical dispersion with the experimental data. We fit the INS data of Cs$_2$CuCl$_4$~\cite{PhysRevB.68.134424} to Eq.~\ref{Eq.teswtenergy} using iterative least squares estimation both by TESWT and $1/S$-LSWT. Our fitting parameters by TESWT are $J=0.480(9)$meV, $J^\prime=0.152(15)$meV and $D=0.012(2)$meV. We give our numerical results in Fig.~\ref{Fig3} and list the different parameters using different method in Table.~\ref{tab:table1}. The parameters of TESWT are with reasonable values. The lose of high order terms may lead our results worse than series expansion results~\cite{JohnPhysRevB.75.174447}. But TESWT is easier to carry out and can give the final ordering vector quickly.

\subsection{Sublattice magnetization}\label{subsec:S}
Next, we study the phase diagram of the anisotropic triangular-lattice. In a spin system, the sublattice magnetization can describe the phase transition behavior. The second-order correction of the sublattice magnetization contributes little to the result. Thus, we only consider the first order correction to the sublattice magnetization as
\begin{equation}
  \Braket{S} = S-\delta S_1 = S-\Braket{a_i^\dag a_i}, \\
\end{equation}
where
\begin{equation}
    \Braket{a_i^\dag a_i} = \Braket{a_\kk^\dag a_\kk} = \Braket{\widetilde{v}_\kk^2}.
\end{equation}
In Fig.~\ref{Fig4} we plot the sublattice magnetization $\langle S \rangle$ variation with spatial anisotropy. Our result without DM interaction is consistent with previous numerical studies~\cite{hauke2011modified,PhysRevB.59.14367}. Consistent with our previous analysis of Fig.~\ref{Fig2}, the spiral order is destroyed at $\alpha\geq 1.2$. In addition, the spiral order is unsafe at $\alpha\leq 0.5$, consistent with modified spin wave results~\cite{hauke2011modified}. The DM interaction, which originates from spin-orbit coupling, helps to generate a non-collinear spin ground state. It is evident from Fig.~\ref{Fig4}, as $\eta$ gets bigger, the phase transformation point in the region $\alpha\leq 0.5$ diminshes until it disappears. On the opposite end, the sublattice magnetization recovers thereby making the $\alpha \geq1.2$ zone less susceptible to drastic effects of quantum fluctuation. These findings suggest that the DM interaction enlarges the region of the spiral state. Our focus in this article is on the multimagnon RIXS spectrum in the spiral phase. Thus, we can use the computed phase diagram to extract the appropriate choice of parameters. We find that TESWT not only gives a consistent physical estimate of the final ordering vector, but also correctly predicts the phase diagram of an anisotropic TLAF, helping to better understand the behavior of the spiral ground state of such a geometrically frustrated system.

%%%%%%%%%%%%%%%%%%%%%%%%%%%%%%%%%%%%%%%%%%%%%%%%%%%%%%%%%%%%%%%%
\section{Indirect RIXS spectra}\label{sec:rixs}
%%%%%%%%%%%%%%%%%%%%%%%%%%%%%%%%%%%%%%%%%%%%%%%%%%%%%%%%%%%%%%%%

\subsection{Noninteracting bi- and trimagnon RIXS}\label{subsec:rixsa}
In this section we calculate the bi- and trimagnon RIXS spectrum. The results in this part use TESWT while the LSWT approach is shown in Appendix~\ref{sec:appendix}.
The indirect RIXS scattering operator, is given by~\cite{EPL.80.47003,PhysRevB.77.134428}
\begin{equation}\label{rixsop}
  \mathcal{R}_\qq=\sum_{i,\bdelta}e^{i\qq\cdot \mathbf{r}_i}[J_{i\bdelta}\bs_i\cdot\bs_{i+\bdelta}-\mathbf{D}_{\bdelta}\cdot\bs_i\times\bs_{i+\bdelta}],
\end{equation}
where $\qq$ is the scattering momentum. In quasiparticle representation, the magnon creation parts of the RIXS scattering operator can be given by
\begin{equation}
\mathcal{R}_\qq=\sum_{1+2=\qq}\widetilde{M}(1,2)b_{1}^\dag b_{2}^\dag+\sum_{1+2+3=\qq}\widetilde{N}(1,2,3)b_{1}^\dag b_{2}^\dag b_{3}^\dag,
\end{equation}
where the bimagnon scattering matrix element is
\begin{eqnarray}\label{2mop-te}
 \widetilde{M}(1,2)=&&\frac{3JS}{2!}\Big\{[\xi_1+\lambda_1+\xi_2+\lambda_2-2(\gamma_\bq+\xi_\qq)](\widetilde{u}_{1}\widetilde{v}_{2}+\widetilde{v}_{1}\widetilde{u}_{2})\non\\
 &&+(\xi_1-\lambda_1+\xi_2-\lambda_2)(\widetilde{u}_{1}\widetilde{u}_{2}+\widetilde{v}_{1}\widetilde{v}_{2})\Big\},
\end{eqnarray}
and the trimagnon scattering matrix element is
\begin{eqnarray}\label{3mop-te}
  \widetilde{N}(1,2,3)=&&\frac{3JS}{3!}i\sqrt{\frac{3}{2SN}}\big[(\bar{\gamma}_1-\bar{\gamma}_{2+3}+\frac{1}{4}\bar{\gamma}_\qq)(\widetilde{u}_1+\widetilde{v}_1)\non\\
  &&\times (\widetilde{u}_2\widetilde{v}_3+\widetilde{v}_2\widetilde{u}_3)+(\bar{\gamma}_2-\bar{\gamma}_{1+3}+\frac{1}{4}\bar{\gamma}_\qq)(\widetilde{u}_2+\widetilde{v}_2)\non\\
  &&\times (\widetilde{u}_1\widetilde{v}_3+\widetilde{v}_1\widetilde{u}_3)+(\bar{\gamma}_3-\bar{\gamma}_{1+2}+\frac{1}{4}\bar{\gamma}_\qq)(\widetilde{u}_3+\widetilde{v}_3)\non\\
  &&\times (\widetilde{u}_1\widetilde{v}_2+\widetilde{v}_1\widetilde{u}_2)\big].
\end{eqnarray}
We neglect the corrections from magnon interactions for the trimagnon intensity, which appear at $1/S^{2}$ order. Next, using Eqs.~\eqref{fullrixs1} and ~\eqref{fullrixs2} stated in Appendix~\ref{sec:appendix} we obtain the following expressions for $I_2(\qq,\omega)$ (noninteracting bimagnon) and $I_3(\qq,\omega)$ (trimagnon) scattering intensity
\begin{eqnarray}\label{Eq:bare2m-te}
I_2(\qq,\omega)&=&2\sum_{\kk}\widetilde{M}_{\kk+\qq,-\kk}^2\delta(\omega-\omega_{\kk+\qq}^{(0)}-\omega_{\kk}^{(0)}),\\ \label{Eq:bare3m}
I_3(\qq,\omega)&=&6\sum_{\kk,\pp}\widetilde{N}_{\kk,\qq-\kk-\pp,\pp}^2\delta(\omega-\omega_{\kk}^{(0)}-\omega_{\qq-\kk-\pp}^{(0)}-\omega_{\pp}^{(0)}),
\end{eqnarray}
where $\omega_\kk^{(0)}=\widetilde{\varepsilon}_\kk+\varepsilon_\kk^c$.

\begin{figure}
\centering\includegraphics[width=3.4in]{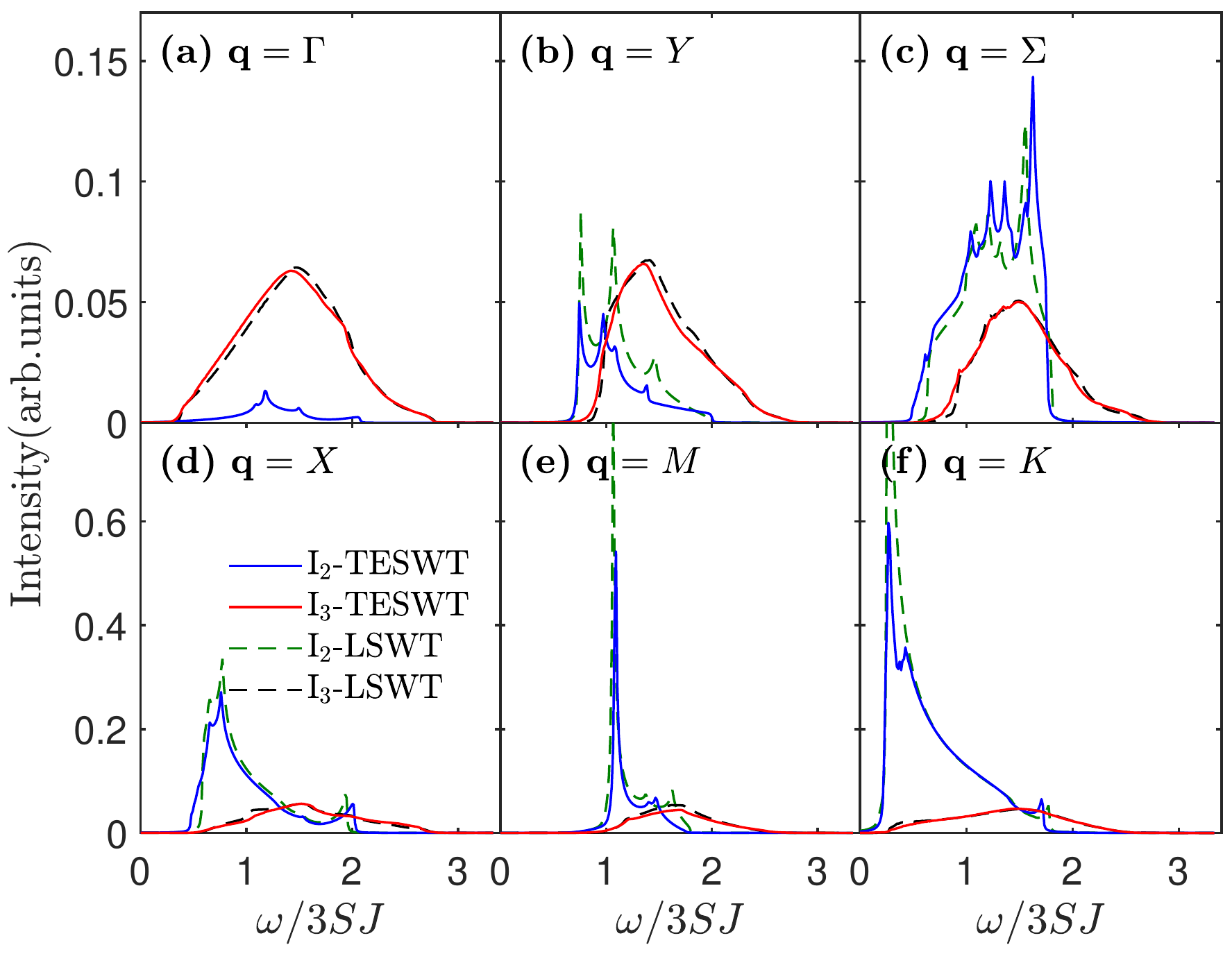}
\caption{Noninteracting bimagnon spectra across high symmetry BZ points. The line plots compare results from TESWT against LSWT for $\alpha=0.8$ and $\eta=0$.}
\label{Fig5}
\end{figure}

In Fig.~\ref{Fig5} we display our results of the noninteracting bi- and trimagnon RIXS spectra at various points across the BZ. Overall the agreement between the LSWT and the TESWT formalism is reasonable. Our TESWT result generates more peaks for the bimagnon intensity. We note that in the isotropic regime $\alpha = 1$, our TESWT results are identical with the LSWT formalism since the final ordering vector {\bf Q} equals the classical vector {\bf Q}$_{cl}$, see Fig.~\ref{Fig11}. As discussed earlier, the TESWT is the physically correct formalism in the presence of anisotropy.
%%%%%%%%%%%%%%%%%%%%%%%%%%%%%%%%%%%%%%%%%%%%%%%%%%%%%%%%%%
%%%%%%%%%%%%%%%%%%%%%%%%%%%%%%%%%%%%%%%%%%%%%%%%%%%%%%%%%%

\subsection{Interacting bimagnon RIXS spectra}\label{subsec:rixsb}
We now proceed with the analysis of $1/S$ correction to the two-magnon Green's function by taking into account both the self-energy correction
to the single magnon propagator $G$ according to the Dyson equation and the vertex insertions to the two-particle propagator $\Pi$
which satisfies the Bethe-Salpeter (BS) equation~\cite{PhysRevB.45.7127,PhysRevB.92.035109}%~\cite{PhysRevB.4.992,PhysRevB.45.7127}.

%%%%%%%%%%%%%%%%%%%%%%%%%%%%%%%%
\begin{figure}
\centering\includegraphics[scale=0.4]{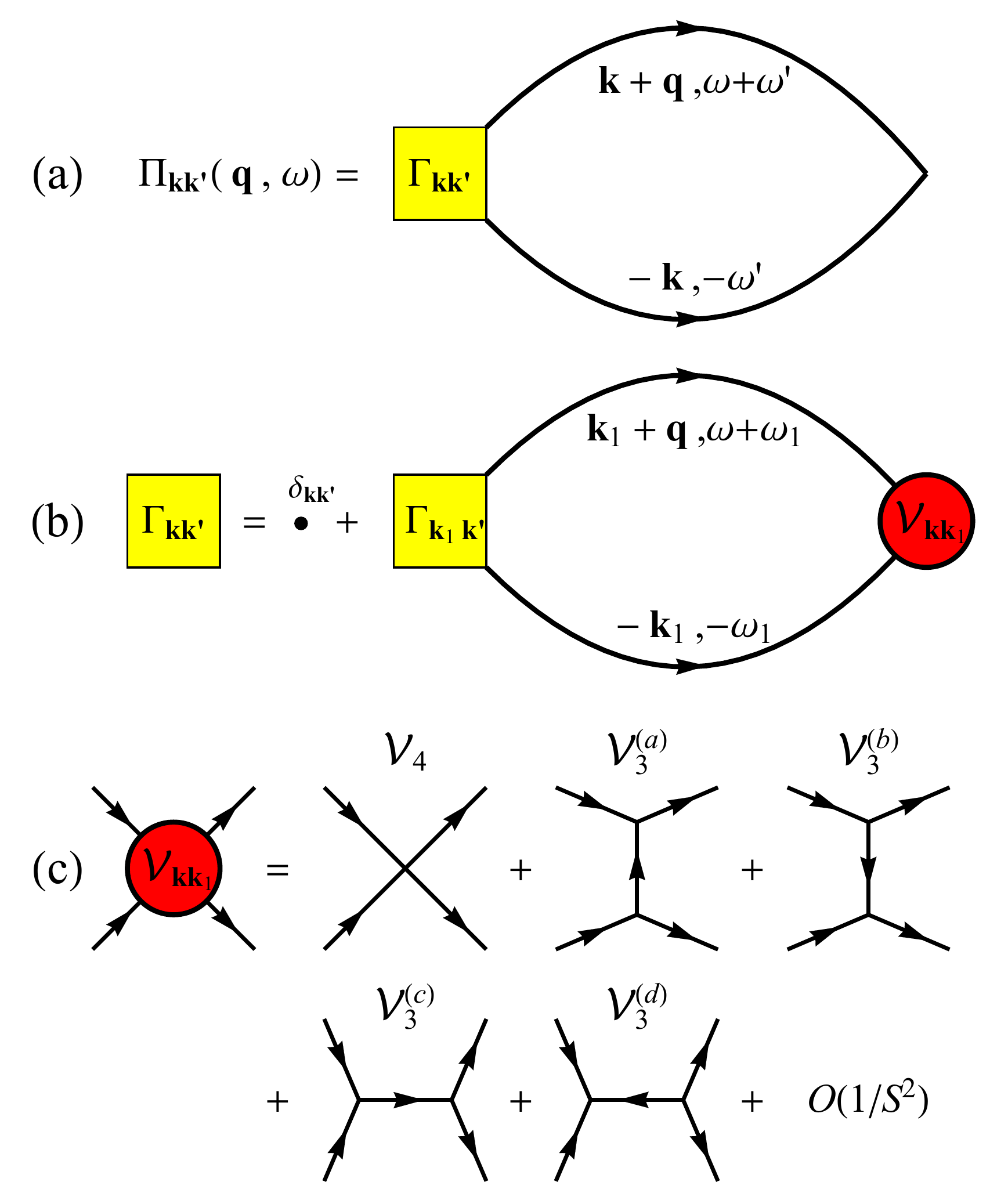}
\caption{Diagrammatic representation for the two-magnon interactions.
(a) Two-magnon propagator $\Pi_{\kk\kk'}(\qq,\omega)$,
(b) Bethe-Salpeter euqation of the vertex function $\Gamma_{\kk\kk'}(\omega,\omega')$ and
(c) the $1/S$ order irreducible interaction $\mathcal{V}_{\mathrm{IR}}$.
Solid lines with an arrow in (a) and (b) stand for the single-magnon propagators. The total irreducible bimagnon scattering vertices can be classified into -
direct ($\mathcal{V}_4$) and indirect ($\mathcal{V}^{a-d}_3$) contributions. Note, the direct ladder interaction leads to a stable magnon interaction event, but the indirect collision process has contributions from virtual decays and recombination.}
\label{Fig6}
\end{figure}
%The direct collision between the two main magnons is caused by the quartic vertex $\mathcal{V}_4$ while the cubic vertices $\mathcal{V}^{a-d}_3$ represent the indirect magnon$-$magnon interactions. Note that in the direct ladder interaction events the two main magnons created in the RIXS process are stable while virtual decays and recombination are allowed in the indirect collision process.
%%%%%%%%%%%%%%%%%%%%%%%%%%%%%%%%
Using the procedure outlined in our prior work \cite{PhysRevB.92.035109} and Feynman rules in momentum space, we obtain the following equations for the two$-$particle propagator and the associated vertex function as
\begin{align}\label{Eq:BSeq1}
\Pi_{\kk\kk'}(\qq,\omega)=&2i\int\frac{\mathrm{d}\omega'}{2\pi}\mathrm{G}_{\kk+\qq}(\omega+\omega')
\mathrm{G}_{-\kk}(-\omega')\Gamma_{\kk\kk'}(\omega,\omega'),\\ \label{Eq:BSeq2}
\Gamma_{\kk\kk'}(\omega,\omega')=&\delta_{\kk\kk'}+\sum_{\kk_1}2i\int\frac{\mathrm{d}\omega_1}{2\pi}
\mathrm{G}_{\kk_1+\qq}(\omega+\omega_1)\mathrm{G}_{-\kk_1}(-\omega_1)\nonumber\\
&\times\mathcal{V}^{\mathrm{IR}}_{\kk\kk_1}(\omega^\prime,\omega_1)\Gamma_{\kk_1\kk'}(\omega,\omega_1),
\end{align}
where the basic one-magnon propagator up to $1/S$ order is now given by
\begin{eqnarray}
 \mathrm{G}^{-1}(\kk,\omega)&=&\omega-\omega_\kk^{(0)}+i0^+.
\end{eqnarray}
%The factor of $2$ in Eqs.(\ref{Eq:BSeq1}) and (\ref{Eq:BSeq2}) stems from
%the two sets of contributions differing by the interchange of dummy momenta $\kk_{(1)}+\qq$ and $-\kk_{(1)}$ according to the Wick's theorem.
The lowest order two-particle irreducible interaction vertex in Fig.~\ref{Fig6}(c) reads
\begin{eqnarray}
\mathcal{V}_{\mathrm{IR}}
=\mathcal{V}_4+\mathcal{V}_3^{(a)}+\mathcal{V}_3^{(b)}+\mathcal{V}_3^{(c)}+\mathcal{V}_3^{(d)},
\end{eqnarray}
in which the frequency-independent four-point vertex $\mathcal{V}_4$ coming from the quartic Hamiltonian can be written as
\begin{eqnarray}
\mathcal{V}_4=\widetilde{V}_c(\kk_1+\qq,-\kk_1;\kk+\qq,-\kk),
\end{eqnarray}
and the other four vertices $\mathcal{V}_3^{(a-d)}$ in the same $1/S$ order which are assembled from two three-point vertices and one frequency-dependent
propagator can be written as
\begin{eqnarray}
\mathcal{V}_3^{(a)}=&&\frac{1}{(2!)^2}[\widetilde{V}_{a}(\kk_1+\qq,\kk-\kk_1;\kk+\qq)\mathrm{G}_0(\kk-\kk_1,\omega'-\omega_1)\non\\
&&\times \widetilde{V}^\ast_{a}(-\kk,\kk-\kk_1;-\kk_1)],\\
\mathcal{V}_3^{(b)}=&&\frac{1}{(2!)^2}[\widetilde{V}^\ast_{a}(\kk+\qq,\kk_1-\kk;\kk_1+\qq)\mathrm{G}_0(\kk_1-\kk,\omega_1-\omega')\non\\
&&\times \widetilde{V}_{a}(-\kk_1,\kk_1-\kk;-\kk)],\\
\mathcal{V}_3^{(c)}=&&\frac{1}{(2!)^2}[\widetilde{V}_{a}(\kk_1+\qq,-\kk_1;\qq)\mathrm{G}_0(\qq,\omega)\non\\
&&\times \widetilde{V}^\ast_{a}(\kk+\qq,-\kk;\qq)],\\
\mathcal{V}_3^{(d)}=&&\frac{1}{(3!)^2}[\widetilde{V}_{b}(\kk_1+\qq,-\kk_1,-\qq)\mathrm{G}_0(-\qq,-\omega)\non\\
&&\times \widetilde{V}^\ast_{b}(\kk+\qq,-\kk,-\qq)].
\end{eqnarray}
In the above we have retained only the bare propagator $\mathrm{G}_0$ for each intermediate line in $\mathcal{V}_3^{(a-d)}$ in the spirit of $1/S$ expansion. Note, the vertex expressions here are different from those stated within the traditional 1/S-SWT approach \cite{PhysRevB.92.035109}. The vertex expressions here are shifted by the correct TESWT wave vector as represented by the tilde notation. Based on the above generalization, we now derive the final solution of the interacting RIXS intensity from the ladder approximation BS equation.
%We further assume that two on-shell magnons are created and annihilated in the repeated ladder scattering process with $\omega'\approx-\omega_\kk^{(0)}=-\widetilde{\varepsilon}_\kk-\varepsilon_\kk^c$ and $\omega_1\approx-\omega_{\kk_1}^{(0)}=-\widetilde{\varepsilon}_{\kk_1}-\varepsilon_{\kk_1}^c$. This approximation is best for sharp spectral peaks of the two main magnons in the scattering process where all the lowest order irreducible vertices are not explicitly frequency-dependent.
%%%%%%%%%%%%%%%%%%%%%%%%%%%%%%%%%%%%%%%%%%%%
\begin{figure}
\centering\includegraphics[scale=0.35]{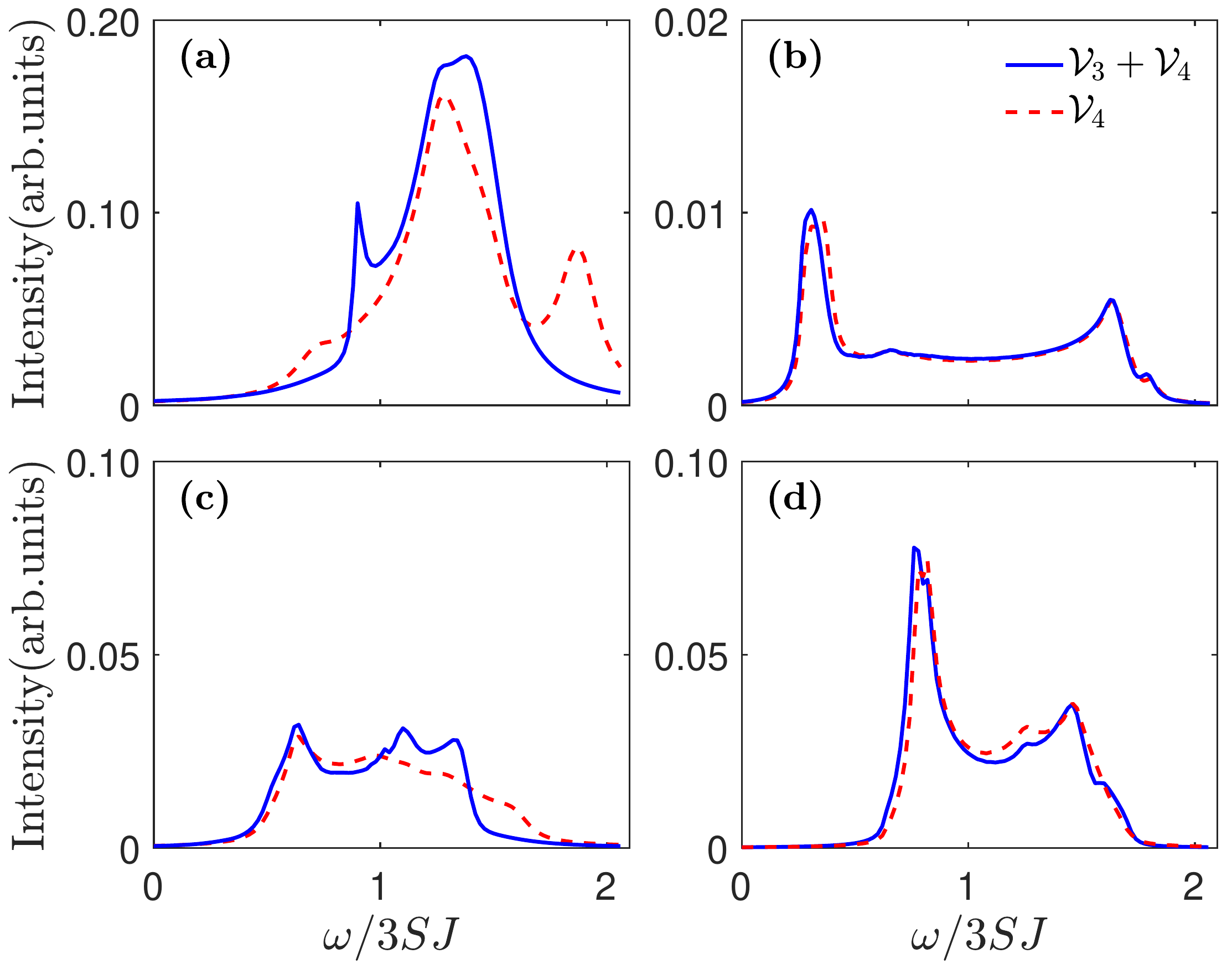}
\caption{Interacting bimagnon RIXS intensity at $\qq=Y(0,\pi/\sqrt{3})$ point with (a) $\alpha=1, \eta=0$, (b) $\alpha=0.316, \eta=0.025$, (c) $\alpha=0.7, \eta=0$ and (d) $\alpha=0.7, \eta=0.05$.
}
\label{Fig7}
\end{figure}
%%%%%%%%%%%%%%%%%%%%%%%%%%%%%%%%%%%%%%%%%%%%%%%%%%%%%%%%%%%%%%%%%%%%%%%%
%We take a numerical approach to compute the RIXS spectrum solving Eq.(\ref{Eq:corrfun}), by summing over $N$ points of $\kk$ in the $1^{st}$ BZ~\cite{PhysRevB.77.174412,PhysRevB.87.174423}.

We adopt a numerical approach to compute the interacting bimagnon RIXS intensity. We assume that two on-shell magnons are created and annihilated in the repeated ladder scattering process
with $\omega'\approx-\omega_\kk^{(0)}=-\widetilde{\varepsilon}_\kk-\varepsilon_\kk^c$ and $\omega_1\approx-\omega_{\kk_1}^{(0)}=-\widetilde{\varepsilon}_{\kk_1}-\varepsilon_{\kk_1}^c$. We substitute (\ref{Eq:BSeq1}) and (\ref{Eq:BSeq2}) into (\ref{Eq:corrfun}) to obtain
\begin{eqnarray}
\chi_2=\sum_{\kk\kk'}\widetilde{M}_{\kk}\widetilde{M}_{\kk'}
\Big[\delta_{\kk\kk^\prime}\Pi_{\kk}
+\Pi_{\kk}\sum_{\kk_1}V_{\kk\kk_1}\Pi_{\kk_1}\Pi_{\kk_1\kk'}\Big],
\end{eqnarray}
where $\Pi_{\kk}=2[\omega-\omega_{\kk+\qq}-\omega_\kk+i0^+]^{-1}$ is the renormalizated two-magnon propagator in the absence of vertex correction. To proceed further we divide the BZ into $N$ points and replace the continuous momenta $(\kk,\kk^\prime,\kk_1)$ with discrete variables $(\mathrm{m,n,l})$. Thus, we can write
\begin{eqnarray}
\mathrm{\hat{\chi}}_{\mathrm{mn}}=\widetilde{M}_\mathrm{m}\widetilde{M}_\mathrm{n}\Big[\delta_\mathrm{mn}\Pi_\mathrm{m}
+\Pi_\mathrm{m}\sum_{\mathrm{l}}V_\mathrm{ml}\Pi_\mathrm{l}\Gamma_\mathrm{ln}\Big].
\end{eqnarray}
%and from Eq.~\ref{Eq:BSeq1}and Eq.~\ref{Eq:BSeq2} we can get
where
\begin{eqnarray}
\Gamma_{mn}=\delta_{mn}+\sum_{l}\Pi_l V_{ml} \Gamma_{ln}.
\end{eqnarray}
Adopting the matrix notation $\Gamma=(\hat{1}-V\Pi)^{-1}$ we obtain the final form of the $\mathrm{\hat{\chi}}$ matrix as
\begin{eqnarray}\label{Eq:fincorr2}
 \mathrm{\hat{\chi}}^T=\mathcal{\hat{\widetilde{D}}}\textbf{[}\mathbf{\hat{1}}-\hat{\Gamma}\textbf{]}^{-1}\mathcal{\hat{\widetilde{G}}},
\end{eqnarray}
where we have defined the following $N\times N$ matrices,
\begin{eqnarray}
\mathbf{\hat{1}}_\mathrm{mn}&=&\delta_\mathrm{mn},\mathcal{\hat{\widetilde{D}}}_\mathrm{mn}=\delta_\mathrm{mn}\widetilde{M}_\mathrm{m},
\\\hat{\Gamma}_\mathrm{mn}&=&\Pi_\mathrm{m}V_\mathrm{mn},
~\mathcal{\hat{\widetilde{G}}}_\mathrm{mn}=\delta_\mathrm{mn}\Pi_\mathrm{m}\widetilde{M}_\mathrm{n}.
\end{eqnarray}
The interacting bimagnon RIXS susceptibility is computed as
\begin{eqnarray}\label{Eq:fincorr}
\mathrm{\chi}_{2}(\qq,\omega)=\sum_{\mathrm{m,n}}\mathrm{\hat{\chi}}_{\mathrm{mn}}.
\end{eqnarray}
%%%%%%%%%%%%%%%%%%%%%%%%%%%%%%%%
\begin{figure}
\centering\includegraphics[width=3.4in]{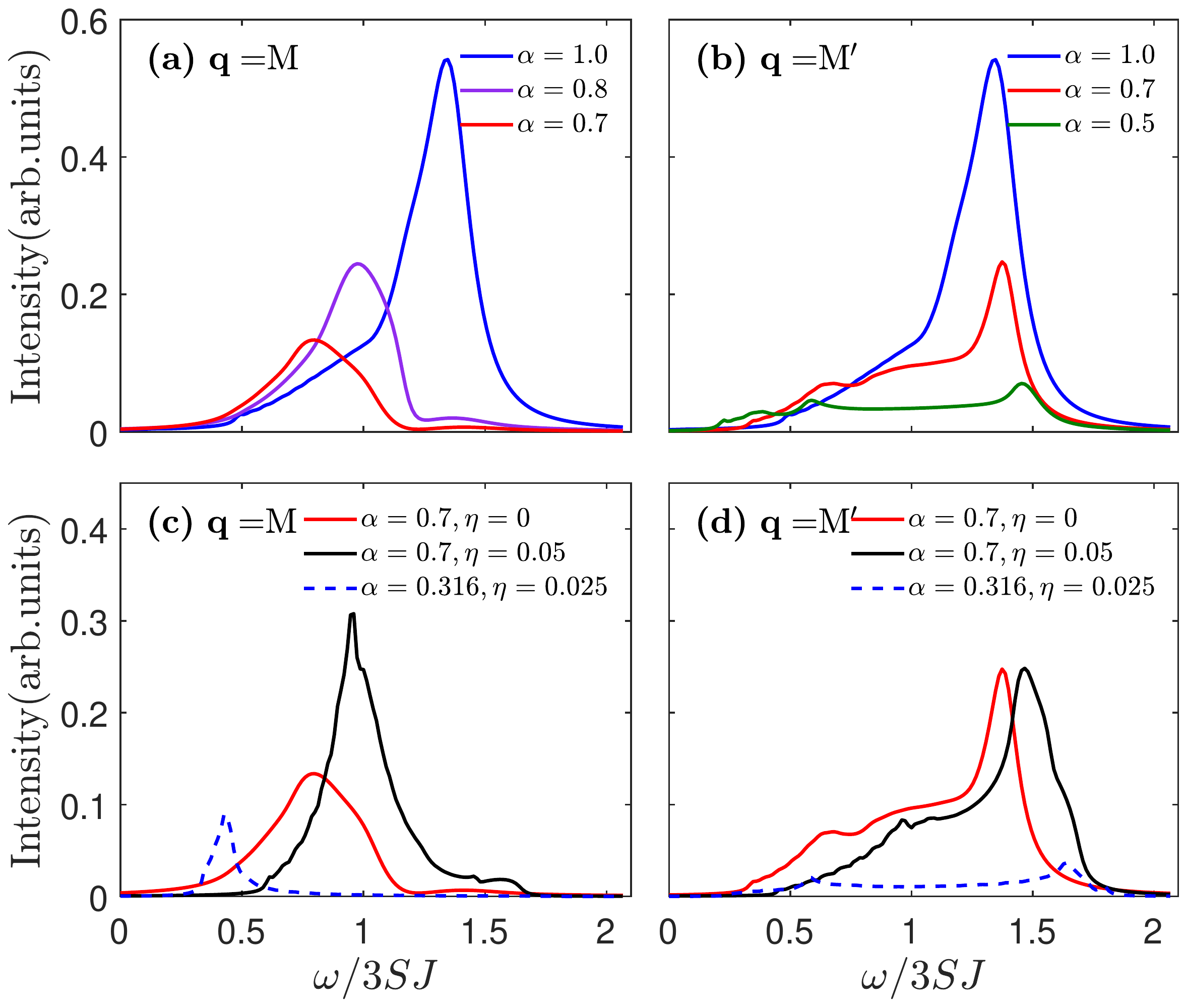}
\caption{Influence of spatial anisotropy and DM interaction on the interacting bimagnon intensity at the two inequivalent roton points
M$(0,2\pi/\sqrt{3})$ and M$^\prime(\pi,\pi/\sqrt{3})$. The first row shows the effect of spatial anisotropy. The second row is the influence of DM interaction. The dashed line utilizes TESWT fitting parameters for Cs$_2$CuCl$_4$.
}
\label{Fig8}
\end{figure}
%%%%%%%%%%%%%%%%%%%%%%%%%%%%%%%%
We use Eqs.~\eqref{Eq:BSeq1} - \eqref{Eq:fincorr} and Eq.~\eqref{fullrixs1} stated in Appendix~\ref{sec:appendix} to numercially compute our interacting bimagnon RIXS intensity at $M$, $M^{\prime}$, and $Y$ BZ points.

In Fig.~\ref{Fig7} we show the spectra at the $Y$ point. The first panel is a reproduction of our previous result reported in Ref.~\onlinecite{PhysRevB.92.035109}. In Fig.~\ref{Fig7}(b) we display the result of TESWT Cs$_2$CuCl$_4$ RIXS. Compared to the isotropic case or to the other anisotropic situations, panels (c) and (d), this spectrum is substantially broadened. With enhanced anisotropy the lattice can be envisioned as disintegrating into a set of loosely coupled chains. Thus, instead of bimagnons one can expect the emergence of spinons as is expected in 1d systems.  1d RIXS has been able to capture multi-spinon excitations \cite{PhysRevLett.103.047401,Schlappa2018}. Thus, the predicted RIXS spectrum feature could be used to confirm quasi-1d to 2d dimensional crossover features of Cs$_2$CuCl$_4$~\cite{PhysRevB.74.180403}.  In Figs.~\ref{Fig7}(c) or \ref{Fig7}(d) we can compare the effects of including a tiny DM interaction. We find that there is a prominent low energy peak with a relatively muted higher energy response. This tiny DM interaction does not bring about any spectral down- or up- shift. The spectral weight is simply redistributed.

\subsection{RIXS signatures at roton points}\label{subsec:rixsc}
%%%%%%%%%%%%%%%%%%%%%%%%%%%%%%%%%%%%%%%%%%%%%%%%%%%%%%%%
\begin{figure*}
\centering\includegraphics[width=6.8in]{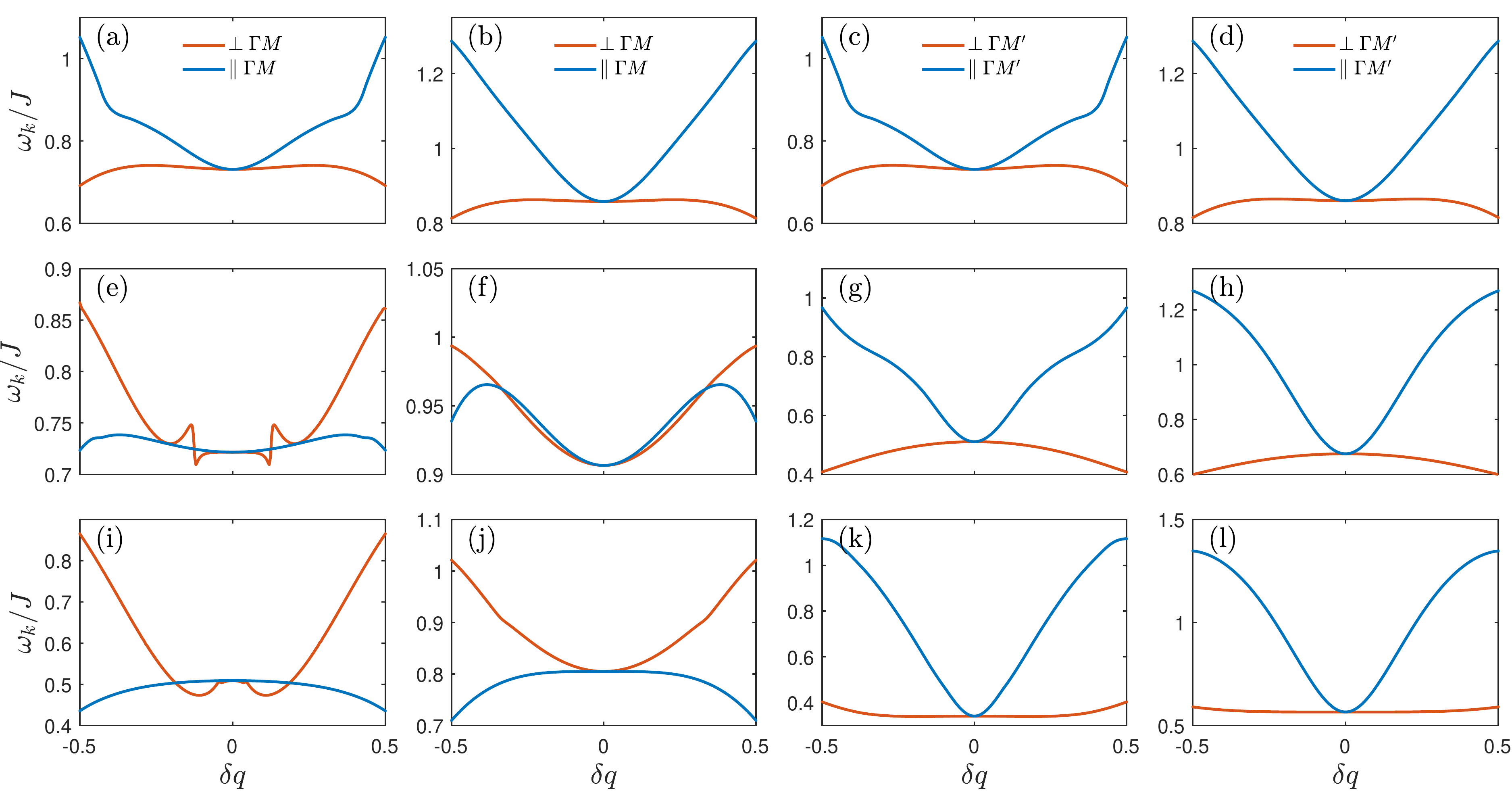}
\caption{The evolution of the roton minimum at $M$ (the first two columns) and $M^\prime$ (the last two columns) points for the $S=\frac{1}{2}$ spiral antiferromagnet on the triangular lattice with (the first and third column) $\eta=0$ and (the second and fourth column) $\eta=0.05$. (a)-(d) $\alpha=1$, (e)-(h) $\alpha=0.7$, (i)-(l) $\alpha=0.5$. The abscissa is defined as $\Delta Q=\frac{2\pi}{3}\delta q$ for parallel to $\Gamma M$ and $\Delta Q=\frac{2\pi}{\sqrt 3}\delta q$ for perpendicular to $\Gamma M$, so as $\Gamma M^\prime$.
}
\label{Fig9}
\end{figure*}
%%%%%%%%%%%%%%%%%%%%%%%%%%%%%%%%%%%%%%%%%%%%%%%%%%%%%%%%

\begin{centering}
%%%%%%%%%%%%%%%%%%%%%%%%%%%%%%%%%%%%%%%%%%%%%%%%%%%%%%%%%%
\begin{figure*}[t]
\centering\includegraphics[scale=0.48]{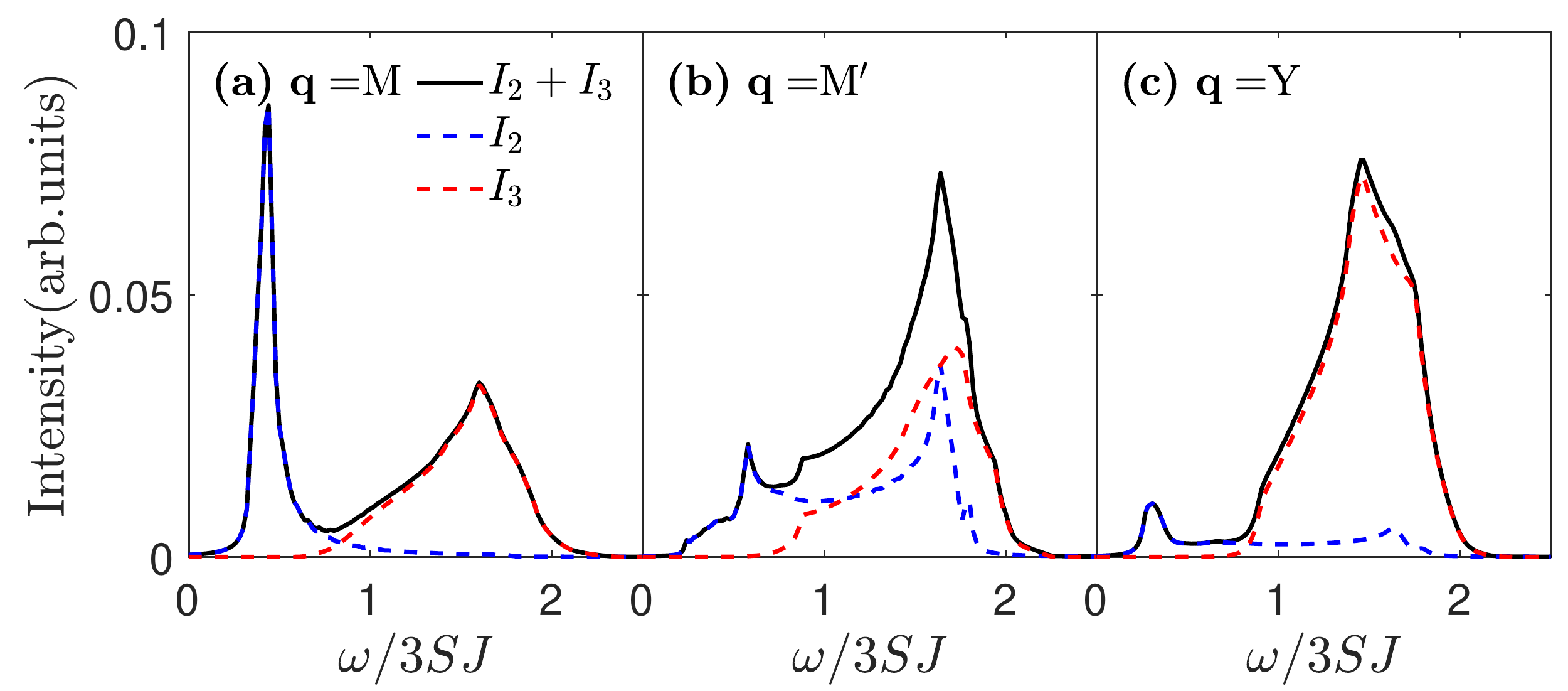}
\caption{Total indirect RIXS spectra $I_2+I_3$ (black solid line) at $\qq=M(0,2\pi/\sqrt{3})$, $\qq=M^\prime(\pi,\pi/\sqrt{3})$ and $\qq=Y(0,\pi/\sqrt{3})$ with TESWT fitting parameters $\alpha=0.316, \eta=0.025$. The individual interacting bimagnon spectra $I_2$ (blue dashed line) and non-interacting tri-magnon spectra $I_3$ (red dashed line) contributions are shown.
}
\label{Fig10}
\end{figure*}
%%%%%%%%%%%%%%%%%%%%%%%%%%%%%%%%%%%%%%%%%%%%%%%%%%%%%%%%%% Thus, the physical explanation of Fig.~\ref{Fig8} could be qualitatively explained by the dispersion in Fig.~\ref{Fig9}.
\end{centering}
In Fig.~\ref{Fig8} we display the interacting RIXS intensity variation at the two anisotropic roton points $\qq=M$ and $\qq=M^\prime$ with varying lattice anisotropy and DM interaction. The anisotropy parameter choices ensure that the TLAF does not decouple into a set of loosely coupled 1d chains, where the bosonization description has been shown to apply ~\cite{PhysRevB.74.180403}. The upper panel Figs.~\ref{Fig8}(a) and \ref{Fig8}(b) are results for zero DM interaction. Note, the two spectrum coincide in the isotropic limit since the two roton points are equivalent due to $C_{3v}$ symmetry of the isotropic triangular lattice~\cite{PhysRevB.92.035109}, while they evolve differently in the presence of spatial anisotropy. In particular, we find that the roton spectra at $\qq=M$ point (the roton point along the $k_y$ direction in BZ) is very sensitive to anisotropy. Though the single-peak structure is stable against $J'/J$, the peak position undergoes a spectral downshift with increased anisotropy. On the other hand, for the $q=M^\prime$ point (along the diagonal BZ direction), the peak location of the spectra does not change much, in comparison to the $M$ point, in the presence of anisotropy. In the lower panel, Figs.~\ref{Fig8}(c) and \ref{Fig8}(d), we display the behavior of the RIXS spectra with DM interaction. Contrary to the isotropic case, the presence of DM interaction introduces a spectral upshift at both $\qq=M$ and $\qq=M^\prime$. The dashed line in the lower panel is the result of using realistic parameters generated from the Cs$_2$CuCl$_4$ INS data fit based on TESWT.

To gain insight into the roton behavior of the RIXS spectra we track the evolution of the roton minimum in the single magnon dispersion along $\Gamma \rightarrow M$ and $M^{\prime}$, both parallel and perpendicular to the BZ path, see Fig.~\ref{Fig9}. A bimagnon excitation requires $\omega_{\kk+\qq}+\omega_{\kk}$ amount energy. We notice that the one magnon dispersion along $M$ displays more sensitivity compared to that along $M^{\prime}$. The asymmetrical sensitivity to the dispersion stiffness explains the origins of the differing roton RIXS spectra behavior. Increasing anisotropy reduces the one magnon energy (softening) near the $M$ point (the first column in Fig.~\ref{Fig9}), thus leading to a spectral downshift in Fig.~\ref{Fig8}. Whereas for the $M^\prime$ point, the overall energy scale of the dispersion is not affected (the third column in Fig.~\ref{Fig9}). We observe neither a drastic hardening nor softening. Thus, the RIXS spectrum holds steady without any shift. The softening and subsequent flattening of the dispersion at the $M$ point suggests that for the anisotropic TLAF, the roton feature is retained more at the $M$ point compared to the $M^{\prime}$. However, inclusions of the DM interaction increases the one magnon energy both near $M$ and $M^\prime$ points (the second and fourth column in Fig.~\ref{Fig9}), introducing a spectral upshift. This could be understood by the fact that DM interaction introduces a gap, thus it requires more energy to create a single magnon and in turn a bimagnon excitation.

The evolution of the spectral height in Fig.~\ref{Fig8} can also be explained. As anisotropy weakens the coupling between the TLAF spins to transform the material to a quasi-1d spin chain, it is more difficult to create a bimagnon excitation. In RIXS, this will cause a decrease in the value of the bimagnon scattering matrix element $|\widetilde{M}(\kk+\qq,-\kk)|$ in turn leading to a reduction in the spectral weight, see Fig.~\ref{Fig8}(a) and \ref{Fig8}(b). On the contrary, the presence of the DM interaction encourages interactions beyond the traditional Heisenberg type. Thus, it assists with the creation of bimagnons, see Fig.\ref{Fig8}(c), where the spectral weight increases. But for the ${\bf q}=M^\prime$ point, the actual nature of the magnon bands is not affected by the DM interaction, see Fig.~\ref{Fig9} fourth column. Thus, the height of the RIXS spectrum does not change with DM interaction. Note, in all the above discussion we have assumed that the triangular lattice does not break down to a set of coupled 1d spin chains.  The $\alpha=0.5$ RIXS spectra could well describe the Cs$_2$CuBr$_4$ compound.

\subsection{Total RIXS}\label{subsec:totrixs}

In Fig.~\ref{Fig10} we report the total RIXS spectrum for Cs$_2$CuCl$_4$ with TESWT fitting parameters. The total RIXS spectrum comprises of the bi- and trimangon response. We use Eqs.~\eqref{fullrixs1} and ~\eqref{fullrixs2} to compute the spectrum. The interacting bimagnon (Eq.~\eqref{Eq:fincorr}) and noninteracting trimagnon intensity (Eq.\eqref{Eq:bare3m}) are summed over to get the total RIXS spectrum. As expected, the trimagnon peak is located at a higher energy than the bimagnon response. In the response for the $M$ and $Y$ points, the main peaks are separated, see Figs.~\ref{Fig10}(a) and \ref{Fig10}(c). At the $M^\prime$ point in Fig.~\ref{Fig10}(b), a small bigmagnon peak is obvious while the main peaks of bi- and trimagnon are mixed.

We note that the spectrum height of the bimagnon undergoes a special evolution. Bimagnon has a height near the boundary of BZ ($M$ and $M\prime$ points) but vanishes when it is close to the center of BZ ($Y$ point). A similar trend for the bigmanon can also be observed in Figs.~\ref{Fig5} and \ref{Fig11}. This is due to the behavior of the RIXS scattering element from the indirect $K$ -edge RIXS scattering operator in Eq.~\eqref{rixsop}. For wave vector choice $\qq$ close to the high symmetry $\Gamma$ point, the RIXS bimagnon matrix element occuring from $\mathcal{R}_\qq$ gives a vanishingly small contribution. Thus the spectral weight of the bimagnon is substantially weakened near the $\Gamma$ point. Without DM interaction, the contribution is purely from the trimagnon excitations at the $\Gamma$ point in the isotropic TLAF, see Fig.~\ref{Fig11}(a). The above observations on the total RIXS spectrum should be helpful in distinguishing the contributions of the two different multimagnon excitations.

%%%%%%%%%%%%%%%%%%%%%%%%%%%%%%%%%%%%%%%%%%%%%%%%%%%%%%%%%%%%%%%%
\section{Conclusion}\label{sec:Conclu}
%%%%%%%%%%%%%%%%%%%%%%%%%%%%%%%%%%%%%%%%%%%%%%%%%%%%%%%%%%%%%%%%
Due to the possible realization of various unusual ordered or disordered phases, frustrated magnetism is an active area of research in condensed matter physics~\cite{StarykhRPP.78.052502}. Traditionally, information on the magnetic ground state and single magnon excitations is inferred from inelastic neutron scattering (INS) experiments \cite{PhysRevLett.87.037202,PhysRevLett.86.1335}. However, with the advent of RIXS spectroscopy experimentalists now have a probe that can comprehensively investigate a wide range of energy and momentum values in BZ.

In this article, we have demonstrated the application of a recently proposed spin-wave theory scheme called TESWT to the indirect $K$ -edge RIXS. As highlighted in this paper it is not a trivial matter to ensure that the sanctity of the spin spiral state is preserved. We performed a TESWT fitting of Cs$_2$CuCl$_4$ INS data, which gives $\alpha\approx0.316$ and $\eta\approx 0.025$. Using these realistic parameters we computed the indirect $K$ -edge bi- and trimagnon RIXS spectra within TESWT formalism. Our results allow us to confirm that in contrast to the isotropic model, quantum fluctuations in the noncollinear anisotropic TLAF can generate divergent fluctuations with drastic effects on the magnetic phase diagram. We find that the behavior of the RIXS spectra is influenced with the occurence of two inequivalent rotonlike points, $M(0,2 \pi/\sqrt{3})$ and $M^{\prime}(\pi,\pi/\sqrt{3})$. While the roton RIXS spectra at the $M$ point undergoes a spectral downshift with increasing anisotropy, the peak at the $M^\prime$ is not affected. However, the peak at $M^\prime$ does not exhibit any downshift. We believe in the anistorpic case the $M$ point retains more of the roton feature. Finally, we find that in the total RIXS spectra, the features of the bimagnon and the trimagnon are certainly different and thus can be easily distinguished within an experimental setting. {\color{red}While resolution and intensity issues may plague the $K$ -edge}, we hope the calculation in this paper and our past publication \cite{PhysRevB.92.035109} will inspire experimentalists to improve resolution to test our predicted $K$ -edge RIXS behavior.

In conclusion, our theoretical investigation of the indirect RIXS intensity in the spiral antiferromagnets on the anisotropic triangular lattice demonstrates that RIXS has the potential to probe and provide a comprehensive characterization of the dispersive bimagnon and trimagnon excitations in the TLAF across the entire BZ, which is far beyond the capabilities of traditional low$-$energy optical techniques~\cite{RevModPhys.79.175,JPCM.19.145243,PhysRevB.77.174412,PhysRevB.87.174423}.
%
%{\bf SJ} I think it's okay. There may be two problems. "We calculated the phase diagram with \textbf{various} different parameters". We show only three parameters for phase diagram. Can we say "various"? Another is "the peak at the $M^\prime$ location \textbf{splits}". We didn't mention "split" in Sec.~\ref{subsec:rixsc}. I write a draft there, trying to explain the split.{\bf SJ}
%%%%%%%%%%%%%%%%%%%%%%%%%%%%%%%%%%%%%%%%%%%%%%%%%%%%%%%%%%%%%%%%
%\newpage
\begin{acknowledgments}
We thank Radu Coldea for sharing with us the INS data for Cs$_2$CuCl$_4$. T~D. acknowledges invitation, hospitality, and kind support from Sun Yat-Sen University Grant No. OEMT--2017--KF--06. T. D. acknowledges funding support from Augusta University Scholarly Activity Award. S.~J., C.~L and D.~X.~Y. are supported by NKRDPC Grants No. 2018YFA0306001, No. 2017YFA0206203, NSFC-11574404, NSFG-2015A030313176, National Supercomputer Center in Guangzhou, and Leading Talent Program of Guangdong Special Projects.
\end{acknowledgments}
%\newpage
%%%%%%%%%%%%%%%%%%%%% appendix A %%%%%%%%%%%%%%%%%%%%%%%%%%%%%%%

\appendix
\section{Isotropic TLAF RIXS spectra}\label{sec:appendix}

%%%%%%%%%%%%%%%%%%%%%%%%%%%%%%%%%%%%%%%%%%%%%%%%%%%%%%%%%%
In this Appendix we compare the results of LSWT and TESWT for the isotropic lattice. We apply linear spin wave theory to the calculation of indirect $K$ -edge RIXS spectrum in this section. After the usual HP and Bogoliubov transformation application, the magnon creation parts of the RIXS scattering operator can be expressed as
\begin{equation}
\mathcal{R}_\qq=\sum_{1+2=\qq}M(1,2)b_{1}^\dag b_{2}^\dag+\sum_{1+2+3=\qq}N(1,2,3)b_{1}^\dag b_{2}^\dag b_{3}^\dag,
\end{equation}
where the bimagnon and trimagnon scattering matrix element expression are given by
\begin{eqnarray}\label{2mop}
  M(1,2)=&&\frac{3JS}{2!}\Big\{[\xi_1+\lambda_1+\xi_2+\lambda_2-2(\gamma_\bq+\xi_\qq)](u_{1}v_{2}+v_{1}u_{2})\non\\
  &&+(\xi_1-\lambda_1+\xi_2-\lambda_2)(u_{1}u_{2}+v_{1}v_{2})\Big\},\\ \label{3mop}
  N(1,2,3)=&&\frac{3JS}{3!}i\sqrt{\frac{3}{2SN}}\big[(\bar{\gamma}_1-\bar{\gamma}_{2+3}+\frac{1}{4}\bar{\gamma}_\qq)(u_1+v_1)\non\\
  &&\times (u_2v_3+v_2u_3)+(\bar{\gamma}_2-\bar{\gamma}_{1+3}+\frac{1}{4}\bar{\gamma}_\qq)(u_2+v_2)\non\\
  &&\times (u_1v_3+v_1u_3)+(\bar{\gamma}_3-\bar{\gamma}_{1+2}+\frac{1}{4}\bar{\gamma}_\qq)(u_3+v_3)\non\\
  &&\times (u_1v_2+v_1u_2)\big].
\end{eqnarray}
%%%%%%%%%%%%%%%%%%%%%%%%%%%%%%%%%%%%%%%%%%%%%%%%%%%%%%%%%%
\begin{figure}[b]
\centering\includegraphics[scale=0.48]{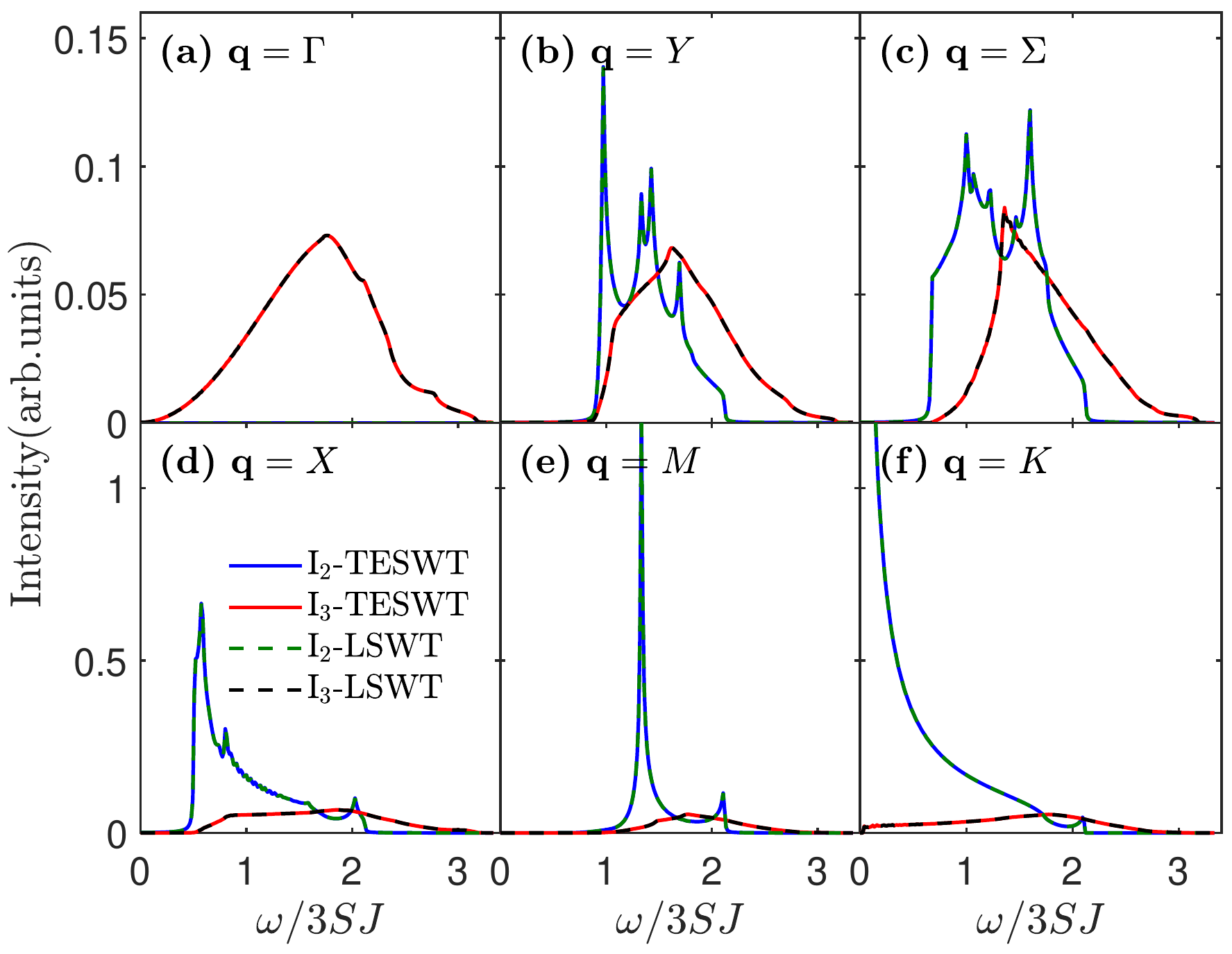}
\caption{Noninteracting bimagnon spectra without spatial anisotropy or DM interaction. The line plots compare results from TESWT against LSWT for $J^\prime/J=1, D/J=0$. The curves of TESWT completely coincide with those of LSWT~\cite{PhysRevB.92.035109}.}
\label{Fig11}
\end{figure}
Note that all the coefficients and functions are defined at the classical ordering vector $\bq_{cl}$ in LSWT. The frequency and momentum dependent magnetic scattering intensity is related to the multimagnon RIXS response function via the fluctuation-dissipation theorem
\begin{equation}\label{fullrixs1}
I(\qq,\omega)=-\frac{1}{\pi}\mathrm{Im}[ \chi_{\mathrm{RIXS}}(\qq,\omega)],
\end{equation}
where the total indirect $K$ -edge RIXS susceptibility is given by
\begin{equation}\label{fullrixs2}
\chi_{\mathrm{RIXS}}(\qq,\omega)=\chi_2(\qq,\omega)+\chi_3(\qq,\omega).
\end{equation}
In the above $\chi_2(\qq,\omega)$ could be either a noninteracting or interacting two$-$magnon susceptibility, but $\chi_3(\qq,\omega)$ is the non$-$interacting three$-$magnon susceptibility.
The susceptibilities can be expressed explicitly from the corresponding multi-magnon Green's function defined as
\begin{eqnarray}\label{Eq:corrfun}
\chi_2(\qq,\omega)&=&\sum_{\kk\kk'}M_{\kk}M_{\kk'}\Pi_{\kk\kk'}(\qq,\omega),\\ \label{Eq:corrfun3}
\chi_3(\qq,\omega)&=&\sum_{\kk\pp;\kk'\pp'}N_{\kk,\pp}N_{\kk',\pp'}\Lambda_{\kk\pp;\kk'\pp'}(\qq,\omega),
\end{eqnarray}
where $\Pi$ and $\Lambda$ denote the bi- and trimagnon propagator, respectively. The momentum-dependent two-magnon and three-magnon Green's function in terms of Bogoliubov quasiparticles are defined as
\begin{eqnarray}\label{Eq:2mgreen}
i\Pi_{\kk\kk'}(\qq,t)&=&\langle\mathcal{T}b_{\kk+\qq}(t)b_{-\kk}(t)b_{\kk^\prime+\qq}^\dag b_{-\kk^\prime}^\dag\rangle,\\ \label{Eq:3mgreen}
i\Lambda_{\kk\pp;\kk'\pp'}(\qq,t)&=&\langle\mathcal{T}b_{\kk}(t)b_{\qq-\kk-\pp}(t)b_{\pp}(t)b_{\kk^\prime}^\dag b_{\qq-\kk^\prime-\pp^\prime}^\dag b_{\pp^\prime}^\dag \rangle,
\end{eqnarray}
where $\mathcal{T}$ is the time-ordering operator and $\langle\cdot\rangle$ is the average of the ground state.
Using Eq.~(\ref{Eq:2mgreen}) and Eq.~(\ref{Eq:3mgreen}), we can compute the noninteracting and the interacting RIXS spectra. The non-interacting spectrum can be calculated by applying Wick's theorem to Eq.~(\ref{Eq:2mgreen}) and Eq.~(\ref{Eq:3mgreen}). The final expressions are stated in Eqs.~\eqref{Eq:bare2m-te} and ~\eqref{Eq:bare3m}.

%
%We obtain the folloexpressions for the noninteracting bimagnon and trimagnon scattering intensity
%\begin{eqnarray}\label{Eq:bare2m}
%I_2(\qq,\omega)&=&2\sum_{\kk}M_{\kk+\qq,-\kk}^2\delta(\omega-\varepsilon_{\kk+\qq}-\varepsilon_{\kk}),\\ \label{Eq:bare3m}
%I_3(\qq,\omega)&=&6\sum_{\kk,\pp}N_{\kk,\qq-\kk-\pp,\pp}^2\delta(\omega-\varepsilon_{\kk}-\varepsilon_{\qq-\kk-\pp}-\varepsilon_{\pp}).
%\end{eqnarray}
%The LSWT results are shown in Fig.~\ref{Fig5} and Fig.~\ref{Fig10}. We then show the reults of interacting bimagnon spectra for LSWT in Fig.~\ref{Fig7} and Fig.~\ref{Fig7}, the process is similar with Sec.~\ref{subsec:rixsb}.

%\bibliographystyle{apsrev4-1}
\bibliography{reftriangular}

\end{document}